\documentclass[12pt,preprint]{aastex}

\usepackage{graphicx}
\usepackage{natbib}
\usepackage{epsfig}

\newcommand{\bu}{{\bf u}}

\newcommand{\bg}{{\bf g}}

\newcommand{\be}{{\bf e}}
\newcommand{\grad}{{\mathbf \nabla}}

\renewcommand{\Pr}{{\rm Pr}}

\shorttitle{Double-diffusive convection}
\shortauthors{Mirouh et al.}

\title{A new model for mixing by double-diffusive convection (semi-convection): I. The conditions for
  layer formation}

\author{G. M. Mirouh$^{1,2}$, P. Garaud$^{2}$, S. Stellmach$^3$, A. L. Traxler$^{2,4}$ and T. S. Wood$^{2}$} 
\affil{$^1$ ENS Cachan,  61, avenue Pr\'esident Wilson
94235 Cachan cedex, France \\
$^2$ Department of Applied Mathematics and Statistics, Baskin School of
 Engineering, University of California Santa Cruz, 1156 High Street,
 Santa Cruz, CA 95064, USA \\
$^3$ Institut f\"ur Geophysik, Westf\"alische Wilhelms-Universit\"at
M\"unster, M\"unster D-48149, Germany \\
$^4$ Department of Physics, Florida International University, 11200 SW 8th
Street, Miami, FL 33199, USA}

\begin{abstract}
The process referred to as ``semi-convection'' in astrophysics and 
``double-diffusive convection in the diffusive regime'' in Earth and planetary sciences, 
occurs in stellar and planetary interiors
in regions which are stable according to the Ledoux criterion but
unstable according to the Schwarzschild criterion. In this series of
papers, we analyze the results of an extensive suite of 3D numerical
simulations of the process, and ultimately propose a new 1D prescription for heat and
compositional transport in this regime which can be
used in stellar or planetary structure and evolution models. 
In a preliminary study of the phenomenon, Rosenblum et al. (2011)
showed that, after saturation of the primary instability, a system can 
evolve in one of two possible ways: the induced turbulence either remains 
homogeneous, with very weak transport properties, or transitions into
a thermo-compositional staircase where the transport rate is much
larger (albeit still smaller than in standard convection). 

In this paper, we show that this dichotomous behavior is a
robust property of semi-convection across a wide
region of parameter space. We propose a simple semi-analytical criterion 
to determine whether layer formation is expected or not, and at what
rate it proceeds, as a function
of the background stratification and of the diffusion parameters (viscosity,
thermal diffusivity and compositional diffusivity) only. The
theoretical criterion matches the outcome of our numerical simulations
very adequately in the numerically accessible ``planetary'' parameter
regime, and can easily be extrapolated to the stellar parameter regime. 

Subsequent papers will address more specifically the question of quantifying
transport in the layered case and in the non-layered case. 

\end{abstract}


\begin{document}

\section{Introduction}

\subsection{The physics of semi-convection: double-diffusive convection} 
\label{sec:intro-basic}

The concept of ``semi-convection'', first introduced by
\citet{schwartzchildharm1958}, is often invoked in  
a number of rather different situations \citep{merryfield1995} which nevertheless have one
point in common: they occur in regions which are stable to the Ledoux criterion, but unstable to the Schwarzschild
criterion. Mathematically speaking, this condition can be expressed as 
\begin{eqnarray}
0 < \left( \frac{\partial \ln T}{\partial \ln p} \right) - \left(
  \frac{\partial \ln T}{\partial \ln p} \right)_{\rm ad} & < & \left(
  \frac{\partial \ln \mu}{\partial \ln p} \right) \mbox{   ,
  equivalently    }
0 < \nabla - \nabla_{\rm ad} < \nabla_\mu
\label{eq:sccriterion}
\end{eqnarray}
where $T$, $p$ and $\mu$ are the temperature, gas pressure, and mean
molecular weight, and where the subscript ``ad'' denotes a derivative at
constant entropy. Physically speaking, (\ref{eq:sccriterion}) describes regions
which are thermally unstably stratified, but where standard
convection is suppressed by the presence of significant compositional
gradients. 

The first linear analysis of the stability of ``semi-convective'' regions
was presented by \citet{walin1964} in the oceanographic context, and later by
\citet{kato1966} in the astrophysical context. They both showed that 
a semi-convective region is hydrodynamically unstable, but to a much 
more gentle instability than the one associated with standard 
convection because it relies on doubly-diffusive processes to
grow. It is in fact one of two forms of  so-called ``double-diffusive
convection'' (the other being fingering convection, otherwise known as ``thermohaline convection'' in astrophysics),
and is often referred to as ``double-diffusive convection in the diffusive regime'' in physical oceanography. 
For the sake of clarity and brevity, we will refer to the phenomenon as 
``diffusive convection'' in this series of papers. 

As reviewed by \citet{rosenblum2011}, diffusive convection is principally controlled by three
non-dimensional parameters. The first two characterize the nature of
the fluid considered, and are the Prandtl number Pr and the
diffusivity ratio $\tau$, 
\begin{equation}
{\rm Pr} = \frac{\nu}{\kappa_T} \mbox{    ,   }
\tau = \frac{\kappa_\mu}{\kappa_T} \mbox{    ,   }
\label{eq:Prtaudef}
\end{equation}
where $\nu$, $\kappa_T$ and $\kappa_\mu$ are the viscosity, thermal
diffusivity and compositional diffusivity respectively. For reference,
note that Pr and $\tau$ are very roughly of the order of $10^{-2}$
in giant planet interiors and $10^{-6}$ in
stellar interiors, and that $\tau$ is usually somewhat smaller than $\Pr$. 

The third parameter is the inverse density ratio, defined as
\begin{equation}
R^{-1}_0 = \frac{\nabla_\mu}{\nabla - \nabla_{\rm ad}}\mbox{   , }
\label{eq:R0def}
\end{equation}
which measures the relative importance of the destabilizing thermal stratification
compared with the stabilizing compositional one. A semi-convective region is unstable \citep{walin1964,kato1966} if
\begin{equation}
1 \le R_0^{-1} \le \frac{{\rm Pr} + 1}{{\rm Pr} + \tau}  =
R^{-1}_{\rm c} \mbox{   , }
\label{eq:instabrange}
\end{equation}
whereas regions with $R_0^{-1} < 1$ are unstable to overturning
convection and those with
$R_0^{-1} > R^{-1}_{\rm c}$ are absolutely stable.
Within this parameter range, it can be shown
that the growth rate of the linear modes is complex. Furthermore, in the low Prandtl number 
regime characteristic of stellar and planetary interiors, the real part of the growth rate is 
proportional to the square root of the Prandtl number times the Br\"unt-V\"ais\"al\"a 
frequency (see Appendix A). In the same limit, the typical lengthscale of the 
unstable modes is a thermal diffusion scale. 

Linear stability is unfortunately of rather limited utility, in
particular when it comes to estimating the mixing rates induced by the
diffusive convection. Experiments -- laboratory or numerical -- are the
only way forward.  Since no terrestrial fluid exists with similar values
of Pr and $\tau$, it is tempting to use experimental measurements of
heat and compositional fluxes in different parameter
regimes, in particular laboratory experiments in the heat-salt system relevant for physical
oceanography \citep{lindenshirtcliffe1978}, and extrapolate them to the astrophysical 
case \citep{stevenson1982,Gu04}. However one must be very cautious in
doing so since the Prandtl number of water is typically 4-7, depending
on temperature, whereas the Prandtl number in stellar and planetary
interiors is much lower than one. Since the typical scale of the instability is a
thermal diffusion scale, double-diffusive mixing is a mostly-laminar process at high Pr
and a turbulent one at low Pr. There is no reason to expect that the
laminar scalings should apply to the turbulent case. 

Nevertheless, from a qualitative point of view, one of the most
interesting results from laboratory \citep{turnerstommel1964} and field experiments \citep{timmermans2008} in salt
water is the fact that diffusive convection has a tendency to form 
thermohaline staircases, i.e. well-defined mixed layers separated by
thin, very strongly stratified, and essentially diffusive interfaces. It is often thought that
the necessarily weak transport through these interfaces is
what controls and limits the efficiency of transport by
``layered'' convection. Such
considerations have led \citet{spruit1992} and \citet{Chabrier07c}
for example to propose theories for heat and compositional transport
in astrophysics, which rely on assumptions about the layer heights and
the interface thicknesses. However, it is important to remember that,
until very recently, layered convection had never been demonstrated to exist at low Prandtl
number. As a result, these theories have, by and large, remained un-tested (see
\citet{rosenblum2011} for a review of prior numerical work). 

\subsection{Recent numerical and theoretical results} 
\label{sec:intro-rosen}

\begin{figure}
\includegraphics[width=\textwidth]{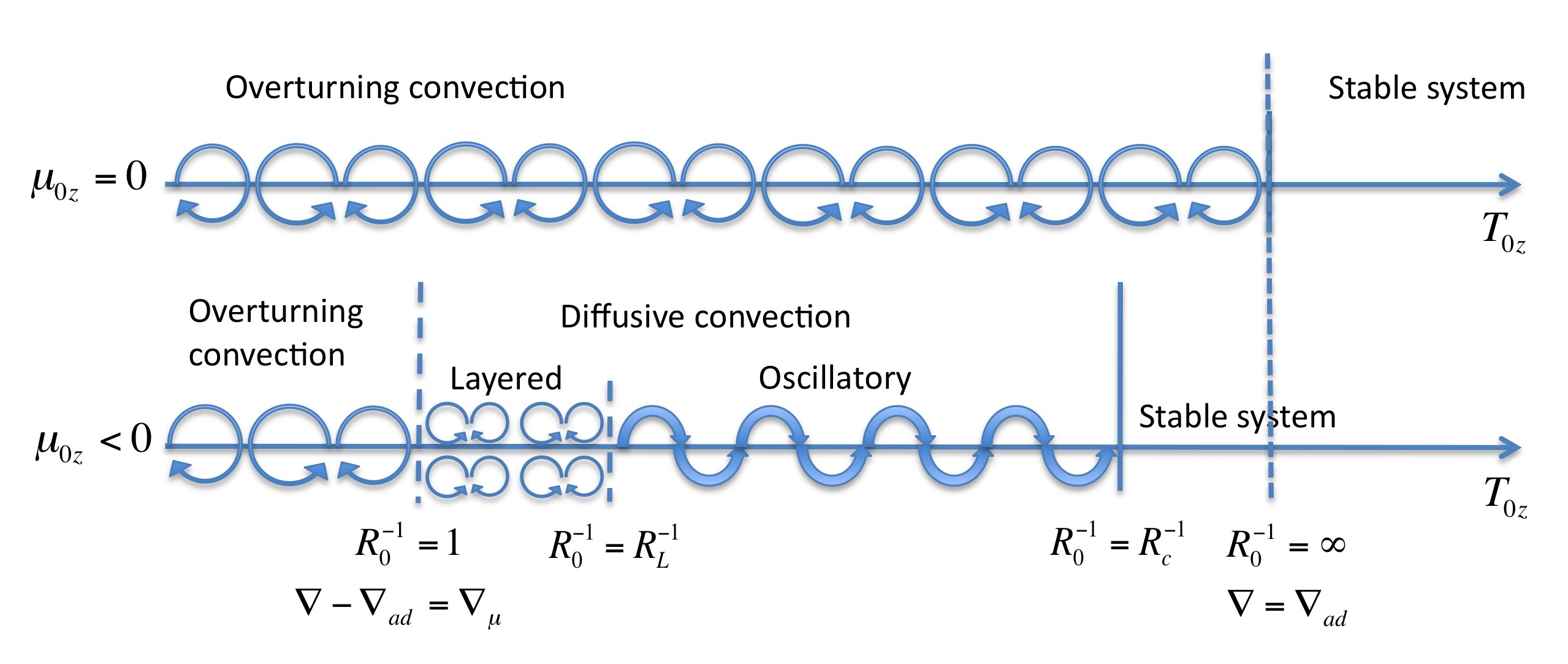}
         \caption{\small Illustration of the various regimes of
           diffusive convection. In systems without
           compositional gradients, the Schwarzschild criterion marks
           the stability boundary between overturning convection and
           absolute stability. In the presence of a stable
           compositional gradient, diffusive convection occurs
           for $R_0^{-1}$ between 1 (which corresponds to the
           Ledoux-stability limit) and $R_c^{-1} =
           (\Pr+1)/(\Pr+\tau)$. Within this range, two possibilities
           arise: for $R_0^{-1} \in [1,R_L^{-1}]$, spontaneous
           transition into 
           layered convection is observed, while for $R_0^{-1} \in
           [R_L^{-1}, R_c^{-1}]$, the system remains in a state
           of weak oscillatory convection. Note that both $R_L^{-1}$ and
         $R_c^{-1}$ depend on $\Pr$ and $\tau$. }
\label{fig:layerregime}
\end{figure}

Recent 3D numerical simulations have finally shed some light on the
subject of transport by diffusive convection. \citet{rosenblum2011} ran an
exhaustive numerical study of the phenomenon for
fixed Prandtl number and diffusivity ratio Pr$=\tau=0.3$, and with
the inverse density ratio $R_0^{-1}$ spanning the entire instability range (\ref{eq:instabrange}).
While still far from any astrophysically-relevant regime, the selected values of Pr
and $\tau$ in these simulations were below unity and therefore in the right ``region'' of
parameter space. \citet{rosenblum2011} found that the instability grows 
as expected and that the early behavior of the fluid can be
satisfactorily explained by considering the 
fastest-growing modes of instability only. However, they discovered that two very
different regimes of diffusive convection are
possible after saturation of
what we will refer to as the ``primary'' instability, depending on the value of the inverse density ratio. The
various regimes are illustrated in Figure \ref{fig:layerregime}.

For large enough $R_0^{-1}$, i.e. for more ``stable'' stratifications, \citet{rosenblum2011} showed that 
the system settles into a homogeneous, 
statistically stationary, weakly turbulent state. The turbulence is
dominated by internal gravity waves\footnote{Recall that the
  background stratification is stably stratified in terms of the
  density and therefore supports standard gravity waves which
  oscillate with the buoyancy frequency.} and mixing occurs principally via
wave-breaking. Heat and compositional
transport are fairly inefficient, and depend sensitively on the inverse density ratio.  
For low enough $R_0^{-1}$ on the other hand, which corresponds to systems closer to the
Ledoux-stability criterion, they observed the spontaneous emergence of thermo-compositional staircases 
after a short adjustment period. These staircases take the form of vigorously
convective layers, which are thermally and compositionally well-mixed,
and separated by fairly sharp interfaces, as observed in the
oceanographic case. The interfaces, however, are far from merely
diffusive and are instead very dynamic, and often pierced by strong
localized updrafts and downdrafts. Later on, the layers are observed to merge, and
each merger is accompanied by a significant increase in the overall 
transport across the staircase. \citet{rosenblum2011}
found that transport in that regime depends sensitively on the mean layer
height, and proposed preliminary scalings to quantify it. The latter 
remain to be verified across a wider region of parameter space.

In fact, insight into the reason for this dichotomous ``layered
vs. non-layered'' convection, can be
gained from recent oceanographic studies of fingering convection, a related
double-diffusive instability of thermally stable fluids
which are destabilized by adverse compositional gradients
\citep{stern1960sfa}. Such conditions are found in the tropical
thermocline for example, where surface
heating and evaporation continually warm up the upper layers of water, and increase
its salt concentration. Crucially, thermohaline staircases are also commonly
found in fingering regions of the ocean
\citep{schmitt2005edm}. \citet{radko2003mlf} studied their
formation in this regime, and showed that they
can naturally emerge as a secondary large-scale ``mean-field'' instability of the
system. More precisely, he showed that horizontally invariant but 
vertically sinusoidal density perturbations grow exponentially out of the homogeneous, small-scale fingering convection, and
eventually overturn into a regularly-spaced staircase. His theory was
later validated by \citet{stellmach2011} via  three-dimensional 
numerical simulations. 

A crucial result of Radko's theory is his identification of a
necessary and sufficient condition for the layer-forming instability, namely that the turbulent flux ratio
$\gamma_{\rm turb}$, defined as the ratio of the turbulent buoyancy flux due to
heat transport, to the turbulent buoyancy flux due to salt transport, 
should be a decreasing function of the density ratio
$R_0$. He thus named the instability ``the $\gamma-$instability''.
The fact that $\gamma_{\rm turb}$ in salt water decreases with $R_0$ for low density ratios, then
increases again for higher density ratios, explains why
thermohaline staircases in the tropical ocean are only found in
regions with low enough $R_0$.  

Recently, \citet{rosenblum2011} showed that Radko's $\gamma-$instability
theory can very easily and naturally be
extended to explain the emergence of staircases in their
own simulations of diffusive convection. The equivalent condition for
instability is that the {\it total} flux ratio $\gamma_{\rm tot}$ (i.e. the
ratio of the total buoyancy fluxes, diffusive plus advective, of heat
to composition respectively), should be a decreasing function of
$R_0$. Since the inverse density ratio $R_0^{-1}$ is a more convenient
parameter in diffusive convection, an equivalent sufficient
condition for instability is that $\gamma_{\rm tot}^{-1}$ should be a
decreasing function of $R_0^{-1}$. For completeness, the $\gamma-$instability
theory is rederived and discussed in Section \ref{sec:newtheory}. \citet{rosenblum2011} found through their systematic exploration of the instability
range that $\gamma_{\rm tot}^{-1}$ has a minimum at about $R_L^{-1} =
1.4$ when  $\Pr=\tau=0.3$. This explains why layers are seen to form for
$R_0^{-1} < 1.4$ in their simulations (at these values of Pr and $\tau$) but not for larger 
$R_0^{-1}$. Furthermore, in the simulations which
do lead to layering, \citet{rosenblum2011}  found that theory and numerical
experiments agree remarkably well on the growth rate of the
$\gamma-$instability. 
Their preliminary study thus suggested that, in order to
know under which conditions layer formation is possible in stars and
giant planets, one simply needs 
to determine if and when $\gamma_{\rm
  tot}^{-1}$ decreases with $R_0^{-1}$, for a given parameter pair $(\Pr, \tau)$.

\subsection{Work outline} 

The findings of \citet{rosenblum2011} lay a very clear path towards
creating a practical model for transport by diffusive convection (semi-convection)
in astrophysics:   
\begin{enumerate}
\item Model the function $\gamma_{\rm tot}^{-1}(R_0^{-1};\Pr,\tau)$ (from
  numerical simulations and/or theoretical calculations) to determine if and when
  layered convection is expected, and verify whether the
  $\gamma$-instability predictions continue to hold at lower $\Pr$ and $\tau$.
\item Characterize transport by layered convection, and in particular,
  its dependence on layer height, $\Pr$, $\tau$ and $R_0^{-1}$.
\item Characterize transport by homogeneous diffusive convection (i.e. in the
  absence of layers).
\end{enumerate}
The first step is addressed in this paper, while steps 2 and 3 are 
deferred to subsequent publications in the series. 

In the present paper, we therefore focus our efforts on a precise
determination of the region of parameter space where layered
convection is expected to occur. We do so using a combination of
numerical simulations and theory.  We discuss the numerical model and
present typical results in Section \ref{sec:model}. We review the
$\gamma-$instability theory in Section \ref{sec:newtheory}.
In Section \ref{sec:gammaextract} we outline
the methodology used for extracting the value of the flux ratio
$\gamma^{-1}_{\rm tot}$ from simulations at numerically-accessible
parameters, and present our results. In Section \ref{sec:gammapred}
we present a simple semi-analytical theory which enables us to
estimate $\gamma^{-1}_{\rm tot}$ for any set of parameters, and
compare it with our  numerical results. In Section \ref{sec:layers}, we show that
the predicted growth rates from the $\gamma-$instability theory match
the results of our numerical simulations very well. Finally, we conclude in Section
\ref{sec:ccl}.

\section{Mathematical model and typical solutions}
\label{sec:model}

\subsection{Mathematical model}
\label{sec:nummodel}
 
As discussed by \citet{rosenblum2011}, the typical lengthscale of the
fastest unstable modes is of the order of meters to
hundreds of meters at most in
the parameter regimes typical of stellar and planetary interiors. They found that the first layers to form are
quite thin, spanning no more than a few of these fastest-growing
wavelengths. This justifies studying diffusive convection (at least,
in the early stages of layer formation and evolution), as a local rather
than a global process.

We consider a local Cartesian domain of size $(L_x,L_y,L_z)$, where 
gravity defines the vertical direction: $\bg = -g \be_z$. The small
domain size permits the use of the Boussinesq approximation \citep{spiegelveronis1960} to the 
governing equations, which are then expressed as 
\begin{eqnarray}
\nabla \cdot  \bu &=& 0 \mbox{   , } \nonumber \\
\frac{\partial T}{\partial t} +  \bu \cdot \nabla
T + (T_{0z} - T^{\rm ad}_{0z}) w &=& \kappa_T \nabla^2 T\mbox{   , }\nonumber \\
\frac{\partial \mu }{\partial t} +  \bu \cdot \nabla
\mu + \mu_{0z} w  &=& \kappa_\mu \nabla^2 \mu \mbox{   , }\nonumber \\
\frac{\partial \bu}{\partial t} + \bu \cdot \nabla
\bu  &=& - \frac{1}{\rho_0} \nabla p
  + \left( \alpha T - \beta \mu  \right)  g \be_z + \nu \nabla^2 \bu
  \mbox{   . }  \label{eq:momdim} 
\end{eqnarray} 
The first of these equations is the continuity equation, where $\bu = (u,v,w)$ is the
velocity field. The temperature and chemical composition fields (the
latter is represented here for example as
the mean molecular weight of the fluid) are expressed as the sum of a
linear background profile ($zT_{0z}$ and $z\mu_{0z}$) plus
triply-periodic perturbations $T$ and $\mu$.
The thermal energy equation 
has been re-written as an advection-diffusion equation for $T$, with
$\kappa_T$ being the thermal diffusivity. The additional term $ - w T_{0z}^{\rm ad} $
is present in compressible fluids but not in incompressible ones, and represents the temperature change due to adiabatic expansion \citep{spiegelveronis1960}. 
Another advection-diffusion equation models the evolution of
the mean-molecular weight perturbation $\mu$, with $\kappa_\mu$ the
corresponding diffusivity. The last equation in (\ref{eq:momdim}) is the momentum equation;
in the Boussinesq approximation, the density perturbation about hydrostatic equilibrium, $\rho$, appears
in the buoyancy term only, and is
linearly related to the temperature and mean molecular weight
perturbations as
\begin{equation}
\frac{\rho}{\rho_0}  = - \alpha T + \beta \mu  \mbox{   , }
\end{equation}
where $\rho_0$ is the (constant) mean density of the region considered, and $\alpha$ and $\beta$ are the coefficients of 
thermal expansion and compositional contraction respectively. The
pressure perturbation is denoted as $p$, and $\nu$ is the
viscosity. All the perturbations satisfy triply-periodic boundary conditions,
\begin{equation}
q(x,y,z,t) = q(x+L_x,y,z,t) =q(x,y+L_y,z,t) = q(x,y,z+L_z,t) \mbox{
  , }
\label{eq:3per}
\end{equation}
where $q \in \{\bu,T,\mu,p\}$. This setup minimizes the effects of boundaries on the system.

Using the following standard non-dimensionalization \citep{rosenblum2011}:
\begin{eqnarray}
&& [l] = d = \left( \frac{\kappa_T \nu}{\alpha g |T_{0z} - T^{\rm ad}_{0z} |} \right)^{1/4} \mbox{    ,   }\nonumber \\
&& [t] = d^2/\kappa_T \mbox{    ,   } \nonumber \\
&& [T] = d |T_{0z} -T^{\rm ad}_{0z}| \mbox{    ,   } \nonumber \\
&& [\mu] = (\alpha/\beta) |T_{0z} -T^{\rm ad}_{0z}|  d  \mbox{    ,   }
\label{eq:nondims}
\end{eqnarray}
the governing equations can be re-written as
\begin{eqnarray}
\frac{1}{{\rm Pr}}\left(\frac{\partial \tilde \bu}{\partial t} +
  \tilde\bu \cdot \nabla  \tilde\bu\right) &=& -\nabla \tilde{p}
  + (\tilde{T}-\tilde{\mu}) \be_z +\nabla^2 \tilde\bu \mbox{   , } \nonumber \\
\frac{\partial \tilde{T}}{\partial t} +  \tilde\bu \cdot \nabla
\tilde{T} -  \tilde w &=& \nabla^2 \tilde{T} \mbox{   , }\nonumber \\
\frac{\partial \tilde{\mu}}{\partial t} +  \tilde\bu \cdot \nabla
\tilde{\mu} - R_0^{-1}  \tilde w  &=& \tau \nabla^2 \tilde{\mu} \mbox{   , }\nonumber \\
\nabla \cdot  \tilde\bu &=& 0 \mbox{   , }
\label{eq:goveqs}
\end{eqnarray} 
where quantities with tildes are dimensionless. 
The three parameters discussed in Section \ref{sec:intro-basic} naturally
appear, namely the Prandtl number Pr, the diffusivity ratio
$\tau$, as well as the inverse density ratio $R_0^{-1}$, see equations
(\ref{eq:Prtaudef}) and (\ref{eq:R0def}). In the notations used here, we also have
\begin{equation}
R_0^{-1} = \frac{\beta \mu_{0z}}{\alpha (T_{0z} - T_{0z}^{\rm ad})} \mbox{   .}
\end{equation}

It is interesting and important to note that this non-dimensional model now
{\it only} knows about the superadiabatic temperature gradient $T_{0z} - T^{\rm ad}_{0z}$ rather than about $T_{0z}$ and $T^{\rm ad}_{0z}$
individually. As such, any two real physical 
systems with the same superadiabaticity, the same density ratio, and the same values of $\Pr$ and $\tau$, will lead to the same non-dimensional set of equations even if their background temperature gradients are different. This degeneracy in the parameters will be discussed in more detail in Section \ref{sec:newtheory}. 

All simulations presented in
this work were obtained using the PADDI code (see \citet{traxler2011b,rosenblum2011}), which solves 
(\ref{eq:goveqs}), in a cubic domain of size $(100d)^3$, subject to boundary conditions
(\ref{eq:3per}) using pseudo-spectral DNS. The selection of the domain size is discussed in Section \ref{sec:gammaextract}.

\subsection{Typical results}
\label{sec:example-res}

\begin{figure}
\includegraphics[width=\textwidth]{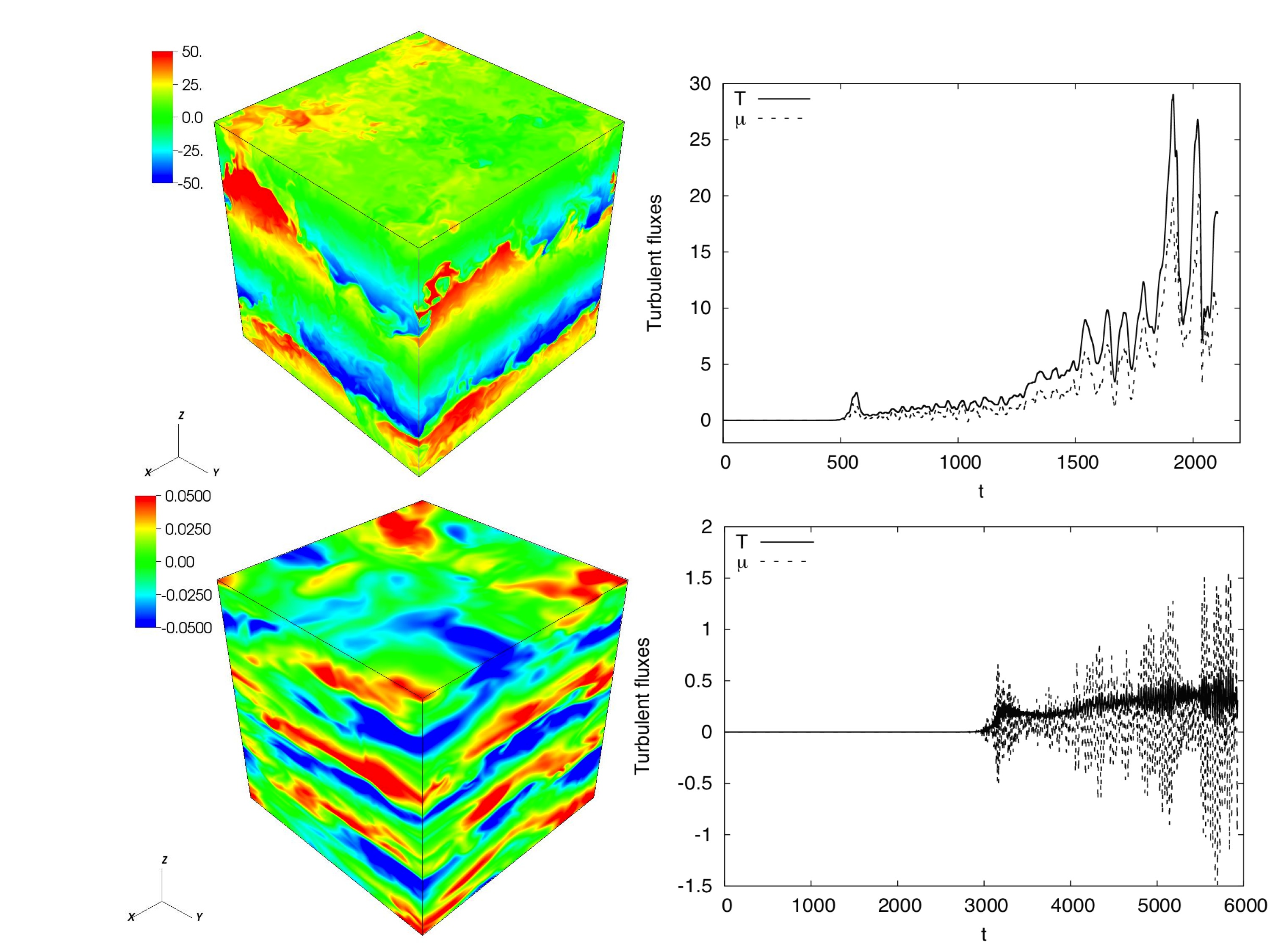}
         \caption{\small Example of simulation results for $\Pr = \tau = 0.03$, for
           $R_0^{-1}=1.5$ (top row) and $R_0^{-1} = 5$ (bottom
           row). The figures on the left are snapshots of the
           compositional perturbation field, at $t=1760$ for the
           $R_0^{-1}=1.5$ case and $t=4600$ for $R_0^{-1}=5$
           case.
           Note the vast difference in the amplitude of the
           perturbations for the two cases: for reference, the total
           compositional contrast across the domain is $\Delta \mu  = 150$ for
           $R_0^{-1}=1.5$ and $\Delta \mu  = 500$ for
           $R_0^{-1}=5$. As a result, the density profile has local inversions in the low $R_0^{-1}$ case (i.e. ``layers''), 
           but remains very close to the background state in the high $R_0^{-1}$ case.
           The figures on the right show the
           corresponding temporal evolution of the non-dimensional
           turbulent fluxes $\langle \tilde{w}\tilde{T} \rangle$ and
           $\langle \tilde{w}\tilde{\mu} \rangle$. Note the stepwise increase in the layered case, with layer formation and each subsequent merger.
 }
\label{fig:kinds}
\end{figure}

Here we present the results of two selected simulations,
which illustrate the behavior of diffusive convection in the planetary
parameter regime. As mentioned in Section \ref{sec:intro-rosen},
previous work at Pr$=\tau=0.3$ showed that the evolution of the system {\it after} saturation of the primary double-diffusive
instability can either result in layer formation or not, depending on
the value of the inverse density ratio $R_0^{-1}$. We confirm that this is still true
 at lower $\Pr$ and $\tau$. This is shown in Figure
 \ref{fig:kinds}, which illustrates the two regimes for $\Pr=\tau =
 0.03$: the layered case, using $R_0^{-1} = 1.5$ (top row), and the
 non-layered case, using $R_0^{-1}  = 5$ (bottom row). In each case, we show on the
left a snapshot of the simulation at a particular time $t$. On the
right we show the temporal evolution of the non-dimensional thermal
and compositional fluxes $\langle \tilde{w} \tilde{T} \rangle$ and $\langle \tilde{w} \tilde{\mu} \rangle$,
where $\langle \cdot \rangle$ denotes 
a spatial average over the entire computational domain.

In the case with low $R_0^{-1}$ (top row), the basic instability rapidly
saturates (around $t=500$ in non-dimensional time units), then
transitions into a layered state, around $t=1200$. Three easily
identifiable layers initially appear, which then merge into two (at
about $t=1500$) then one (at about $t=1800$). The snapshot on
the top left of Figure \ref{fig:kinds} was taken at $t=1760$, and shows the non-dimensional perturbation in
the concentration field in the 2-layered state. The fluxes clearly 
increase in a stepwise manner, first when layers form and then at
each merger. This kind of behavior was already illustrated and
discussed by \citet{rosenblum2011} (see their Figure 7). 

\begin{figure}
\includegraphics[width=\textwidth]{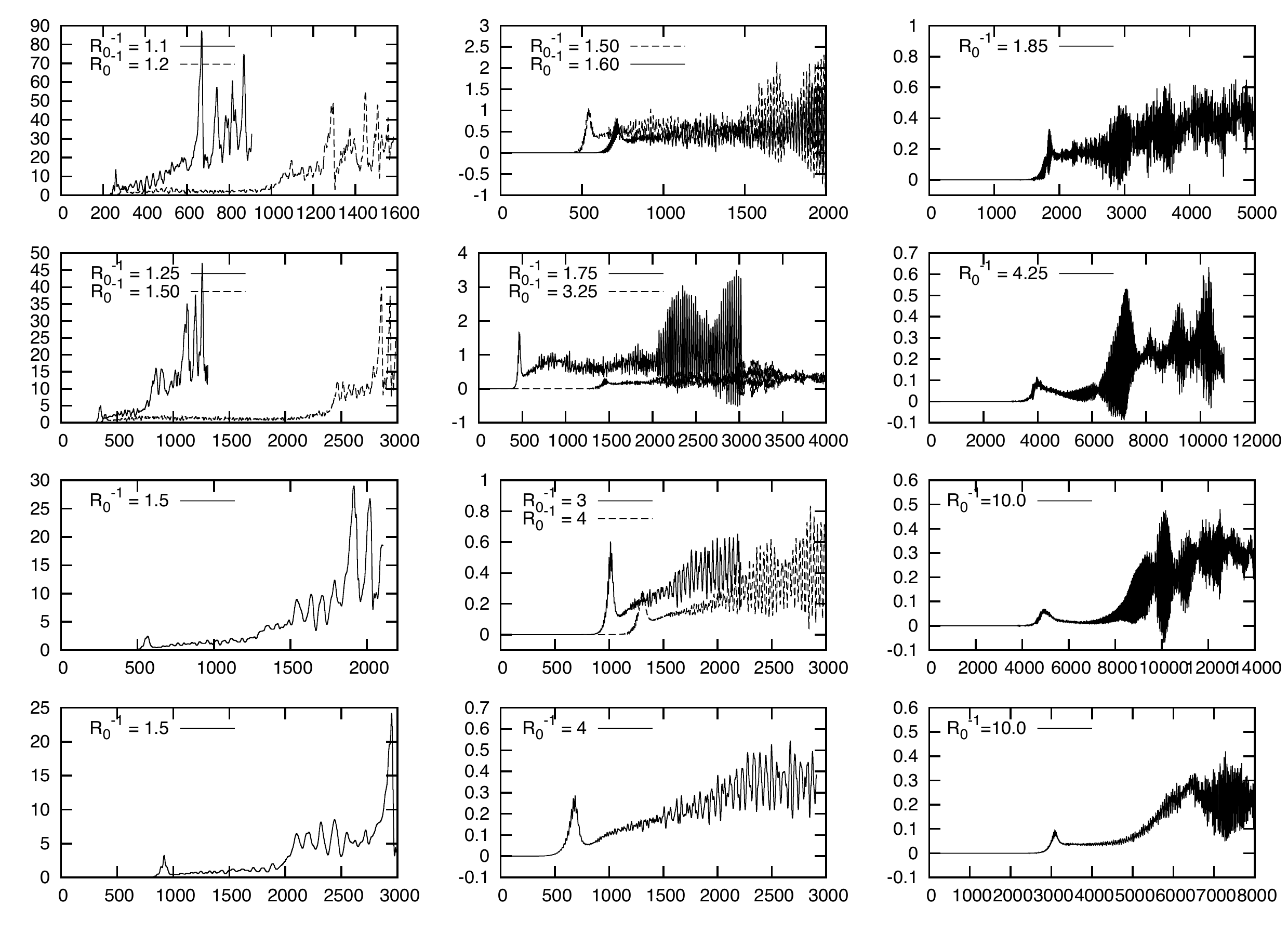}
         \caption{\small Temporal evolution of the turbulent heat flux
           $\langle \tilde{w}\tilde{T} \rangle$ for 
parameter pairs $(\Pr,\tau$) with $\Pr=\tau=$ 0.3, 0.1, 0.03 and 0.01
respectively from top to bottom in each column. In each plot, the
$x-$axis represents the non-dimensional time $t$, and the values of
$R_0^{-1}$ corresponding to each run represented are indicated. The
left-column only shows runs which are found to transition into
layers. The middle and right columns show runs at intermediate and
high values of $R_0^{-1}$ respectively. Runs at the highest values of
$R_0^{-1}$ are often immediately dominated by large-scale gravity waves.}
\label{fig:lotsofruns}
\end{figure}

In the case with high $R_0^{-1}$, the basic instability also grows (although more slowly, as expected from
linear stability) and eventually
saturates around $t=3000$. However, layers never form. Instead, what follows
saturation is what \citet{rosenblum2011} described as being a
homogeneous, weakly convective phase with fairly
inefficient transport properties. 
The snapshot on the bottom left of Figure \ref{fig:kinds} shows the non-dimensional
perturbation in the concentration field at $t= 4600$. Note the small
amplitude of the perturbations, by comparison with the total
compositional contrast across the domain ($\Delta \mu =
500$ here). Upon closer inspection, we find that the
small-scale oscillatory structures that are characteristic of the
homogeneous phase intermittently give way to somewhat larger-scale and more
coherent gravity waves (e.g. here for $4700<t<5200$ and $t>5500$; see
also Figure \ref{fig:lotsofruns}). When this is the case the amplitude
of the wave-induced oscillation in the fluxes
dramatically increases, and the {\it mean} wave-induced
transport also increases, although remains much lower than in the layered
phase. The reason for the emergence of larger-scale waves and their self-organization
remains to be determined, but the phenomenon is fairly ubiquitous at
high $R_0^{-1}$ (see below). This effect will be studied in a subsequent paper.

Figure \ref{fig:lotsofruns} shows the evolution of the turbulent heat flux for 
parameter pairs $(\Pr,\tau$) with $\Pr=\tau$, for selected
$R_0^{-1}$ ranging from values close to overturning instability (left
column), through intermediate values (middle column) to values close to marginal stability (right column). 
The plot clearly illustrates the following trends. Simulations with the lowest
values of $R_0^{-1}$ lead to very rapid layer formation, while those with
slightly larger values of $R_0^{-1}$ can stay in a state of homogeneous
diffusive convection for a very long time before layers emerge 
(see the case of $\Pr = \tau = 0.1$, $R_0^{-1} = 1.5$ for example). At
intermediate values of $R_0^{-1}$, layers never form. The system remains in a state of
homogeneous diffusive convection, and occasionally exhibits intermittent gravity-wave-dominated phases similar to the
one described earlier. Finally, in runs with larger values of $R_0^{-1}$ closer to
marginally stability we always see that the wave-dominated phase
begins very quickly after saturation of the primary instability. 
These various types of behavior need
to be kept in mind when analyzing the data to extract their mean
transport properties, as described in Section \ref{sec:gammaextract}.  

\section{The $\gamma-$instability revisited for case $\grad_{\rm ad}
  \neq 0$}
\label{sec:newtheory}

In this Section we rederive the
$\gamma-$instability theory for the sake of completeness,
and to correct a slight inconsistency in nomenclature discovered in the work of
\citet{rosenblum2011}. While their derivation is technically correct\footnote{See associated Erratum for correction of a typo.}, it requires 
a slight physical re-interpretation of the quantities they define as ``thermal Nusselt number'' and 
``total buoyancy flux ratio'' in order to be fully consistent when $\grad_{\rm ad}
  \neq 0$, as shown below. 

The $\gamma-$instability theory, first proposed by \citet{radko2003mlf} in the context of fingering convection in the ocean, 
is a mean-field theory that describes the development of secondary instabilities in fully-developed 
double-diffusive convection (the theory is in fact valid both in the diffusive regime and in the fingering regime). 
The theory assumes that the system is already in a homogeneous and quasi-steady 
turbulent state and studies its evolution when subject to perturbations on scales much larger
than the turbulent eddies. Accordingly, we begin by averaging the governing non-dimensional equations
(\ref{eq:goveqs}) over all small lengthscales and fast
timescales, and study the evolution of the large-scale, more slowly evolving mean fields. 
 
An emerging staircase is a horizontally invariant structure with no mean flow. If we 
ignore the momentum equation, and neglect mean flows as well as horizontal derivatives, the averaged
non-dimensional thermal and compositional advection-diffusion equations become: 
\begin{eqnarray}
\frac{\partial \overline{T}}{\partial t}  & = & - \frac{\partial F_T^{\rm    tot}}{\partial z} , \nonumber \\
\frac{\partial \overline{\mu}}{\partial t} & = & - \frac{\partial F_\mu^{\rm
    tot}}{\partial z}    \mbox{   , }
\label{eq:gammat1}
\end{eqnarray}
where $\overline{\cdot}$ denotes a spatio-temporal average over
small-scales and short timescales.  Note that 
the vertical fluxes $F_T^{\rm    tot}$ and $F_\mu^{\rm    tot}$ include a diffusive {\it and} a turbulent component.  
The goal is to express them in terms of large-scale fields only and thus close the system of equations, so that the
latter can be solved for the evolution of $\overline{T}(z,t)$ and $\overline{\mu}(z,t)$. 

It is important to note for the upcoming discussion that there is, at this point, some degree of flexibility in the definition of these two fluxes: 
one can add or subtract any constant to $F_T^{\rm tot}$ and $F_\mu^{\rm tot}$  without changing 
the expression $\partial F_{T,\mu}^{\rm tot}/\partial z$. In the original derivation of the $\gamma-$instability, the fluxes are thus taken to be the total non-dimensional heat and compositional fluxes through the system, including the diffusion of the background fields $T_{0z}$ and $\mu_{0z}$.  When expressed non-dimensionally, 
\begin{eqnarray}
F_T^{\rm tot}  = \frac{-\kappa_TT_{0z}}{\kappa_T|T_{0z}-T_{0z}^{\rm ad}|} - \overline{T}_z + \langle \tilde{w} \tilde{T} \rangle \mbox{   ,} \nonumber \\
F_\mu^{\rm tot} = \frac{-\kappa_\mu \mu_{0z}}{\kappa_T (\alpha/\beta)|T_{0z}-T_{0z}^{\rm ad}|}   - \tau \overline{\mu}_z + \langle \tilde{w} \tilde{\mu} \rangle    \mbox{   ,} \nonumber 
\label{eq:wrongftot}
\end{eqnarray}
where the subscript $z$ denotes a derivative with respect to $z$. 


However, while the definition of $F_T^{\rm tot}$ as a total heat flux is intuitive and perfectly adequate for incompressible fluids (for which the theory was originally designed), a subtle but crucial problem emerges for compressible systems, where $T^{\rm ad}_{0z} \neq 0$. The total heat flux explicitly depends on $T_{0z}$, while the original system of equations (\ref{eq:goveqs}) only knows about $T_{0z}-T_{0z}^{\rm ad}$, as discussed in Section \ref{sec:nummodel}. This remark suggests that the dynamically relevant quantity is instead 
\begin{equation}
F_T^{\rm tot}  =  \frac{-\kappa_T (T_{0z}-T_{0z}^{\rm ad})}{\kappa_T |T_{0z}-T_{0z}^{\rm ad}|}  - \overline{T}_z + \langle \tilde{w} \tilde{T} \rangle \mbox{   .}
\end{equation}
The flux thus defined, however, is no longer the total heat flux except when $T^{\rm ad}_{0z} = 0$. 
Simplifying the resulting expressions for $F_T^{\rm tot}$ and $F_\mu^{\rm tot}$ yields
\begin{eqnarray}
F_T^{\rm tot} &=& 1 - \overline{T}_z + \langle \tilde{w} \tilde{T} \rangle  \mbox{   ,} \nonumber \\
F_\mu^{\rm tot} &=&  \tau (R_0^{-1} - \overline{\mu}_z) + \langle \tilde{w} \tilde{\mu} \rangle   \mbox{   ,} 
\label{eq:ftotright}
\end{eqnarray}
These expressions are the ones actually used by \citet{rosenblum2011}. 
The system of equations (\ref{eq:gammat1}) and (\ref{eq:ftotright}) are now mathematically consistent\footnote{An alternative, but equivalent, way to resolve the problem discussed here is to introduce the ``potential temperature'' $\vartheta$ commonly used in the atmospheric literature (e.g. \citet{holton1992}). The evolution equation for $\vartheta$ is identical to that for $T$, except that the adiabatic gradient of $\vartheta$ is zero by construction.} mean-field versions of the original system (\ref{eq:goveqs}).

We now define two non-dimensional quantities:  
\begin{eqnarray}
{\rm Nu}_T  &=& \frac{F_T^{\rm tot}}{ 1 - \overline{T}_z } \nonumber \\
\gamma_{\rm tot}^{-1} &=& \frac{F_\mu^{\rm tot}}{F_T^{\rm tot}}\mbox{   . }
\label{eq:ntdef} 
\end{eqnarray}
The first, ${\rm Nu}_T$, reduces to the much more commonly used temperature thermal Nusselt number (i.e. the ratio of the total heat flux to the diffused heat flux) when $T^{\rm ad}_{0z} = 0$. In what follows, we call it the ``thermal Nusselt number proxy''. 
We also refer to the second, $\gamma_{\rm tot}^{-1}$, as the ``flux ratio'', for simplicity. When $T^{\rm ad}_{0z} = 0$, it reduces
to the total buoyancy flux ratio commonly used in physical oceanography.

The theory then continues exactly as in \citet{rosenblum2011}, by assuming that ${\rm Nu}_T$ and $\gamma_{\rm tot}^{-1}$ each depend only on the fluid parameters $\Pr$ and $\tau$ and on the local inverse density ratio. The latter can vary with $z$ as a result of the large-scale background temperature and compositional perturbations $\overline{T}$ and $\overline{\mu}$, as 
\begin{equation}
R^{-1}_\rho = \frac {\beta ( \mu_{0z} + (\alpha/\beta) |T_{0z}-T^{\rm ad}_{0z}| \overline{\mu}_z) }{\alpha (T_{0z} -T^{\rm ad}_{0z} + |T_{0z} -T^{\rm ad}_{0z}  | \overline{T}_z)} =  \frac {R_0^{-1} -\overline{\mu}_z }{1- \overline{T}_z}    \mbox{   , }
\label{eq:Rrhodef}
\end{equation} 
where, for clarity, we first expressed $R^{-1}_\rho$ as the ratio of dimensional quantities and then as the ratio of non-dimensional quantities.

Combining (\ref{eq:gammat1}), (\ref{eq:ntdef}) and
(\ref{eq:Rrhodef}) yields a nonlinear system of equations describing
the spatio-temporal evolution of the large-scale fields: 
\begin{eqnarray}
\frac{\partial \overline{T}}{\partial t}  & = & - \frac{\partial
  F_T^{\rm tot}  }{\partial z}    \mbox{   , } \nonumber \\
\frac{\partial \overline{\mu}}{\partial t} & = & - \frac{\partial}{\partial z}
\left[  \gamma_{\rm tot}^{-1}(R_\rho^{-1}; \Pr, \tau)  F_T^{\rm tot}  \right]  \mbox{
  , } \nonumber \\
\mbox{  where  } F_T^{\rm tot} &=& {\rm Nu}_T (R_\rho^{-1}; \Pr, \tau) (1 -
\overline{T}_z )  \mbox{    ,     } R^{-1}_\rho =  \frac {R_0^{-1} -\overline{\mu}_z }{1-
  \overline{T}_z} \mbox{   . } 
\label{eq:gammainstab}
\end{eqnarray}
If $\gamma_{\rm tot}^{-1}(R_\rho^{-1}; \Pr, \tau)$ and $ {\rm Nu}_T (R_\rho^{-1}; \Pr, \tau)$ are known, 
finite, non-zero, and smooth enough, then the system of equations is
closed and well-posed. It has a trivial steady-state solution when $\overline{T}_z$ and
$\overline{\mu}_z$ are constant. This solution corresponds to a
homogeneously, diffusively convective state with constant
density ratio $R_\rho^{-1}$. Without loss of generality, we can choose
our reference state $T_{0z}, T^{\rm ad}_{0z}$ and $\mu_{0z}$ to be
that steady-state solution, in which case $R_\rho^{-1} = R_0^{-1}$,
and $\overline{T}_z = \overline{\mu}_z = 0$. The flux ratio and thermal Nusselt number proxy of the homogeneous
background turbulent state are noted as ${\rm Nu}_0 = {\rm Nu}_T(R_0^{-1})$ and $\gamma_0^{-1} = \gamma_{\rm tot}^{-1}(R_0^{-1})$. 

Solving (\ref{eq:gammainstab}) in the general case is numerically possible if the functions 
${\rm Nu}_T$ and $\gamma_{\rm tot}^{-1}$ are known, but not particularly
informative. However, we can linearize the mean-field equations around
the previously defined homogeneously convective state, assuming that the
large-scale perturbations $\overline{T}$ and $\overline{\mu}$ have
small amplitudes. To linear order, the local inverse density ratio becomes
\begin{equation}
R^{-1}_\rho = R^{-1}_0(1 - R_0 \overline{\mu}_z + \overline{T}_z) \mbox{   . }
\label{R0_linear}
\end{equation}
Noting that ${\rm Nu}_T$ depends on $z$ via $R_\rho^{-1}$, it can be
shown using the chain rule that, to linear order, the temperature equation becomes 
\begin{equation}
\frac{\partial \overline{T}}{\partial t}  = - A_2  ( R_0
\overline{\mu}_{zz} - \overline{T}_{zz})  +  {\rm Nu}_0
\overline{T}_{zz}  \mbox{   ,}
\end{equation}
where 
\begin{equation}
 A_2 = -R^{-1}_0 \left.\frac{d {\rm Nu}_T}{dR^{-1}_\rho}
 \right|_{R^{-1}_0} \mbox{   ,}
\label{eq:A2def}
\end{equation}
while the linearized composition equation is similarly derived to be 
\begin{equation}
\frac{\partial \overline{\mu}}{\partial t} = \gamma_0^{-1} \frac{\partial \overline{T}}{\partial t} - A_1
{\rm Nu}_0  ( R_0 \overline{\mu}_{zz} -
\overline{T}_{zz})  \mbox{   ,}
\end{equation}
where 
\begin{equation}
A_1 = - R^{-1}_0 \left. \frac{d(\gamma^{-1}_{\rm tot})}{dR^{-1}_\rho}
\right|_{R^{-1}_0} \mbox{   .}
\label{eq:A1def}
\end{equation}

Assuming normal modes of the form $ \sim e^{i k  z + \Lambda t}$,
we finally get
\begin{equation}
\Lambda^2 + \Lambda k^2 \left[  A_2 (1 -R_0 \gamma_0^{-1})  +
 {\rm Nu}_0 (1 -  A_1  R_0)  \right]  - k^4 A_1 {\rm Nu}^2_0  R_0
=0  \mbox{   .}
\label{eq:newgammaeq}
\end{equation}
This quadratic recovers the one obtained by \citet{radko2003mlf} and
\citet{rosenblum2011} exactly, the only difference being in the physical interpretation of the quantities ${\rm Nu}_T$ and $\gamma_{\rm tot}^{-1}$, as discussed above, when $\grad_{\rm ad} \neq 0$. 

As originally discussed by \citet{radko2003mlf}, inspection of (\ref{eq:newgammaeq}) shows that the condition for the
existence of growing solutions is that the constant term in the
quadratic should be negative, which only occurs when $A_1$ is
positive, i.e., when $\gamma_{\rm tot}^{-1}$ is a decreasing function
of $R_\rho^{-1}$. In the diffusive case studied here, one can prove by inspection of the sign of the linear
term in (\ref{eq:newgammaeq}) that this sufficient condition 
is also a necessary condition for instability (by showing that even if there are complex conjugate roots to this equation, their real parts are negative). \citet{radko2003mlf} also showed that the $\gamma-$instability theory suffers from an ultraviolet catastrophe whereby 
the mode growth rate is proportional to $k^2$ (so that modes with the
smallest wavelengths always grow most rapidly). The theory, however,
must break down when the layering mode wavelength becomes comparable
with the basic instability wavelength. As a result, the actual mode
which ends up growing out of the homogeneous turbulence is the one
with the smallest wavelength for which the mean-field theory is still
valid. Empirically, we find that the latter typically has a {\it
  vertical} wavelength that is about 2-4 times larger than the {\it
  horizontal} wavelength of the fastest growing mode of the basic
instability according to linear theory (see Appendix A). In other words, the staircase typically forms with an initial step separation of about 25-50$d$. 

Finally, in order to identify more quantitatively the conditions for instability and predict its growth rate, we must measure the turbulent fluxes {\it in the homogeneous phase} of diffusive convection to estimate $\gamma_0^{-1}$, ${\rm Nu}_0$, $A_1$ and $A_2$, for various values of $R_0^{-1}$, $\Pr$ and $\tau$. In all that follows, we therefore limit our definitions of $\gamma_{\rm tot}^{-1}$ and ${\rm Nu}_T$ to the case where $\overline{T} = \overline{\mu} = 0$, and so
\begin{eqnarray}
{\rm Nu}_T  &=& 1 + \langle \tilde{w} \tilde{T} \rangle \mbox{   ,}  \nonumber \\
\gamma_{\rm tot}^{-1} &=&  \frac{\tau R_0^{-1} + \langle \tilde{w} \tilde{\mu} \rangle}{ 1 + \langle \tilde{w} \tilde{T} \rangle }    \mbox{   . }
\label{eq:ntgammadef} 
\end{eqnarray}
We also define for convenience a compositional Nusselt number 
\begin{equation}
{\rm Nu}_\mu = 1 + \frac{\langle \tilde{w} \tilde{\mu} \rangle}{\tau R_0^{-1}} \mbox{   ,} 
\label{eq:numudef}
\end{equation}
which measures the ratio of the total compositional flux to the diffused compositional flux in the homogeneous phase. 
With this definition, 
\begin{equation}
\gamma_{\rm tot}^{-1} = \tau R_0^{-1} \frac{ {\rm Nu}_\mu }{{\rm Nu}_T} \mbox{   .} 
\end{equation}

\section{Measurements of the flux ratio}
\label{sec:gammaextract}

As we have just shown in Section \ref{sec:newtheory}, double-diffusive
layering is expected to occur spontaneously whenever the flux ratio $\gamma^{-1}_{\rm
  tot}$ defined in (\ref{eq:ntgammadef})  is a decreasing function of the inverse density ratio $R^{-1}_0 = \grad_\mu / (\grad - \grad_{\rm ad})$. 
In what follows, we refer to the function $\gamma_{\rm  tot}^{-1}(R_0^{-1})$ as ``the $\gamma-$curve''. 
In order to establish when the $\gamma-$curve decreases and estimate the $\gamma$-instability growth rate, we now perform a series of
numerical experiments, decreasing Pr and $\tau$ down progressively
towards the astrophysically relevant parameter regime, and measure both ${\rm Nu}_T(R_0^{-1};\Pr,\tau)$ and
$\gamma_{\rm  tot}^{-1}(R_0^{-1};\Pr,\tau)$ for the whole range of density ratios
unstable to diffusive convection (see (\ref{eq:instabrange})). Section \ref{sec:expsetup} describes our experimental setup and the manner in which we extract the flux ratio and the Nusselt numbers from the simulations. Section \ref{sec:gammares} presents and discusses our results. 

\subsection{Experimental setup}
\label{sec:expsetup}

As in \citet{rosenblum2011} and \citet{traxler2011a}, we use a computational
box of size $L_x = L_y = L_z = 100d$, which is about 4-6 times the wavelength of the 
fastest-growing mode of instability, regardless of the parameters selected (see
Appendix A). This domain size was found to be
sufficiently large to yield statistically meaningful measurements of the turbulent
fluxes while remaining computationally tractable in the increasingly
extreme parameter regimes studied. 

We consider four values of Pr and $\tau$, equal to 0.3, 0.1, 0.03
and 0.01 respectively. The two smallest values, 0.03 and 0.01, are within the planetary
parameter range. For each (Pr,$\tau$) pair, we run a number of
simulations varying $R_0^{-1}$, selecting preferentially values close
to one to capture the expected decreasing part of the $\gamma$-curve. Since the code is a
Direct Numerical Simulation with no sub-grid model, each numerical experiment has to be
fully resolved on all scales. Prior to each full-scale run, we test
various resolutions and select the most appropriate one based on
inspection of the vorticity, velocity and chemical
composition field profiles and spectra. Runs with $R_0^{-1}$ close to unity 
require the highest spatial resolution while runs with $R_0^{-1}$
close to marginal stability require lower spatial resolution, but much
higher temporal resolution and longer integration times to follow
simultaneously the buoyancy frequency timescale and the much slower
instability growth and saturation timescales. 
Tables 1 and 2 summarize the
parameters selected and resolution for all our simulations.

\begin{table}
\begin{center}
\begin{tabular}{cccccc}
\\
\tableline
Pr  &$ \tau $&$ R_0^{-1} $&$ N_{x,y},N_z $&$ t_{\rm tot} $& layers? \\
\hline
\hline
$   0.3 $&$   0.3 $&$  1.10 $&$ 384,384 $&$   907 $&Y      \\
$   0.3 $&$   0.3 $&$  1.15 $&$ 384,192 $&$  1424 $&Y      \\
$   0.3 $&$   0.3 $&$  1.20 $&$ 192,192 $&$  1582 $&Y      \\
$   0.3 $&$   0.3 $&$  1.25 $&$ 192,192 $&$  3251 $&Y      \\
$   0.3 $&$   0.3 $&$  1.35 $&$ 192,192 $&$  2570 $&?      \\
$   0.3 $&$   0.3 $&$  1.50 $&$  96,96 $&$  1999 $&N   \\
$   0.3 $&$   0.3 $&$  1.60 $&$  96,96 $&$  1999 $&N      \\
$   0.3 $&$   0.3 $&$  1.85 $&$  96,96 $&$ 15000 $&N      \\
\hline
$   0.1 $&$   0.1 $&$  1.10 $&$ 384,384 $&$  1114 $&Y      \\
$   0.1 $&$   0.1 $&$  1.25 $&$ 384,384 $&$  1310 $&Y      \\
$   0.1 $&$   0.1 $&$  1.50 $&$ 192,192 $&$  3095 $&Y      \\
$   0.1 $&$   0.1 $&$  1.75 $&$ 192,192 $&$  3028 $&?      \\
$   0.1 $&$   0.1 $&$  2.25 $&$ 192,192 $&$  3531 $&N      \\
$   0.1 $&$   0.1 $&$  3.25 $&$ 192,192 $&$  4138 $&N      \\
$   0.1 $&$   0.1 $&$  4.25 $&$ 192,192 $&$ 10870 $&N      \\
$   0.1 $&$   0.1 $&$  5.00 $&$ 192,192 $&$ 22146 $&N      \\
\hline
$   0.03 $&$   0.03 $&$  1.50 $&$ 576,768 $&$  2104 $&Y      \\
$   0.03 $&$   0.03 $&$  2.00 $&$ 576,576 $&$  1587 $&?     \\
$   0.03 $&$   0.03 $&$  2.50 $&$ 576,576 $&$  1311 $&?      \\
$   0.03 $&$   0.03 $&$  3.00 $&$ 576,576 $&$  2215 $&N      \\
$   0.03 $&$   0.03 $&$  4.00 $&$ 384,384 $&$  3148 $&N      \\
$   0.03 $&$   0.03 $&$  5.00 $&$ 288,288 $&$  5929 $&N      \\
$   0.03 $&$   0.03 $&$ 10.00 $&$ 192,192 $&$ 14845 $&N      \\
\hline
$   0.01 $&$   0.01 $&$  1.50 $&$ 576,576 $&$  2987 $&Y      \\
$   0.01 $&$   0.01 $&$  2.00 $&$ 576,576 $&$  4163 $&?      \\
$   0.01 $&$   0.01 $&$  2.50 $&$ 576,576 $&$  1745 $&?      \\
$   0.01 $&$   0.01 $&$  3.00 $&$ 576,576 $&$  2114 $&N     \\
$   0.01 $&$   0.01 $&$  4.00 $&$ 384,384 $&$  2911 $&N      \\
$   0.01 $&$   0.01 $&$ 10.00 $&$ 288,288 $&$  8138 $&N      \\
\tableline
\end{tabular}
\caption{Presentation of the various runs performed. The first three
  columns present the system parameters. All runs are in cubic domain
  of size $(100d)^3$. The resolution (in terms of equivalent
  mesh-points $N_{x,y},N_z$) is always the same for the two horizontal direction, but
occasionally differs in the vertical direction for runs that are
expected to transition into
layers. The total integration time is given in non-dimensional units
as $t_{\rm tot}$. Finally, we indicate whether we see layers emerge or
not. Runs with a question mark are runs for which we might expect layer
formation based on the $\gamma-$instability criterion, and the actual
position of the minimum of the curve, but where we
have not seen evidence for it (yet). }
 \end{center}
\label{run-summary}
\end{table}

\begin{table}
\begin{center}
\begin{tabular}{cccccc}
\\
\tableline
Pr  &$ \tau $&$ R_0^{-1} $&$ N_{x,y},N_z $&$ t_{\rm tot} $& layers? \\
\hline
\hline
$   0.3 $&$   0.1 $&$  1.10 $&$ 384,384 $&$   761 $&Y      \\
$   0.3 $&$   0.1 $&$  1.20 $&$ 240,240 $&$   787 $&Y      \\
$   0.3 $&$   0.1 $&$  1.40 $&$ 192,192 $&$  1316 $&Y    \\
$   0.3 $&$   0.1 $&$  1.70 $&$ 192,192 $&$  1960 $&?      \\
$   0.3 $&$   0.1 $&$  2.00 $&$ 192,192 $&$  1472 $&N      \\
$   0.3 $&$   0.1 $&$  3.00 $&$ 192,192 $&$  5506 $&N      \\
\hline
$   0.1 $&$   0.3 $&$  1.10 $&$ 192,192 $&$  1759 $&Y      \\
$   0.1 $&$   0.3 $&$  1.20 $&$ 240,240 $&$  1923 $&Y      \\
$   0.1 $&$   0.3 $&$  1.30 $&$ 192,192 $&$  1702 $&?     \\
$   0.1 $&$   0.3 $&$  1.50 $&$ 192,192 $&$  1946 $&N      \\
$   0.1 $&$   0.3 $&$  2.00 $&$ 192,192 $&$  4314 $&N      \\
\hline
$   0.3 $&$   0.03 $&$  1.10 $&$ 576,576 $&$   430 $&Y      \\
$   0.3 $&$   0.03 $&$  1.25 $&$ 384,384 $&$   628 $&Y      \\
$   0.3 $&$   0.03 $&$  1.50 $&$ 384,384 $&$  1052 $&Y     \\
$   0.3 $&$   0.03 $&$  2.00 $&$ 288,288 $&$   937 $& ?      \\
$   0.3 $&$   0.03 $&$  3.00 $&$ 192,192 $&$  6847 $&N      \\
\hline
$   0.03 $&$   0.30 $&$  1.10 $&$ 576,576 $&$  1574 $&Y      \\
$   0.03 $&$   0.30 $&$  1.20 $&$ 384,384 $&$  2262 $&Y      \\
$   0.03 $&$   0.30 $&$  1.35 $&$ 384,384 $&$  4100 $&?     \\
$   0.03 $&$   0.30 $&$  1.50 $&$ 384,384 $&$  3177 $&N      \\
$   0.03 $&$   0.30 $&$  2.00 $&$ 288,288 $&$  5750 $&N      \\
\tableline
\end{tabular}
\caption{(Continued from Table 1.) }
 \end{center}
\end{table}


For each simulation, the PADDI code returns the non-dimensional {\it instantaneous} fluxes
integrated over the entire computational domain as diagnostics of the
simulations. However, the thermal Nusselt number proxy and flux ratio defined in the derivation
of the $\gamma-$instability theory (see Section \ref{sec:newtheory})
are only meaningful when viewed as temporal averages taken during a time where the
system is in the assumed homogeneous, quasi-steady, diffusively convective state. 
Identifying that state, unfortunately, turns out to be significantly more difficult than expected. 
Figure \ref{fig:lotsofruns} shows that
transport in diffusive convection is much more 
variable than in the related fingering regime, where the layering
theory and the methods for extracting small-scale fluxes were first
derived \citep{traxler2011a}. The underlying reason for this difference actually remains to be determined. 
Our selected domain size, for example, was initially chosen by analogy with studies of
transport in fingering convection by \citet{traxler2011a} and
\citet{traxler2011b}, where it was found to be 
``[...] small enough to suppress any secondary large-scale
instabilities'' \citep{traxler2011a}. We find here, by contrast, that
large-scale perturbations (layers, large-scale gravity waves) do in
fact grow even in such a small domain, and cause the observed variability in the fluxes. 

In Appendix B, we discuss the problem in detail, and propose a systematic method to
identify the homogeneous state described above, and 
extract the fluxes, Nusselt numbers and flux ratio in that phase. The results are presented below. 

\subsection{Nusselt numbers and flux ratio}
\label{sec:gammares}

Our measurements for ${\rm Nu}_T$ and
${\rm N}_\mu$ and 
$\gamma_{\rm tot}^{-1} $, obtained using the method described in Appendix B, 
are summarized in Figures \ref{fig:nusr} and \ref{fig:gamma} respectively
for each parameter pair (${\rm Pr},\tau$). The full dataset is
presented in Appendix B. 

\begin{figure}[h!]
\includegraphics[width=0.5\textwidth]{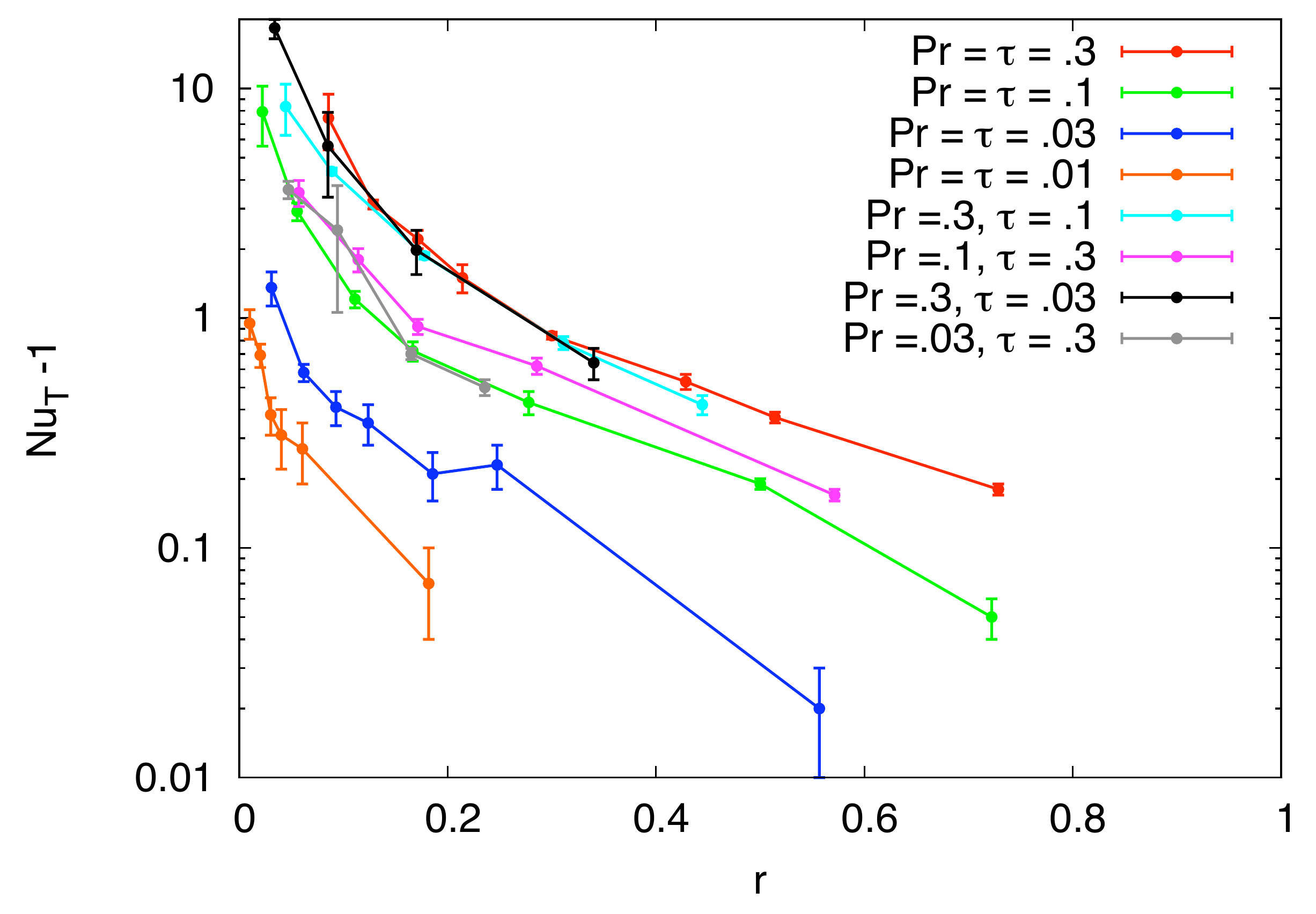}\includegraphics[width=0.5\textwidth]{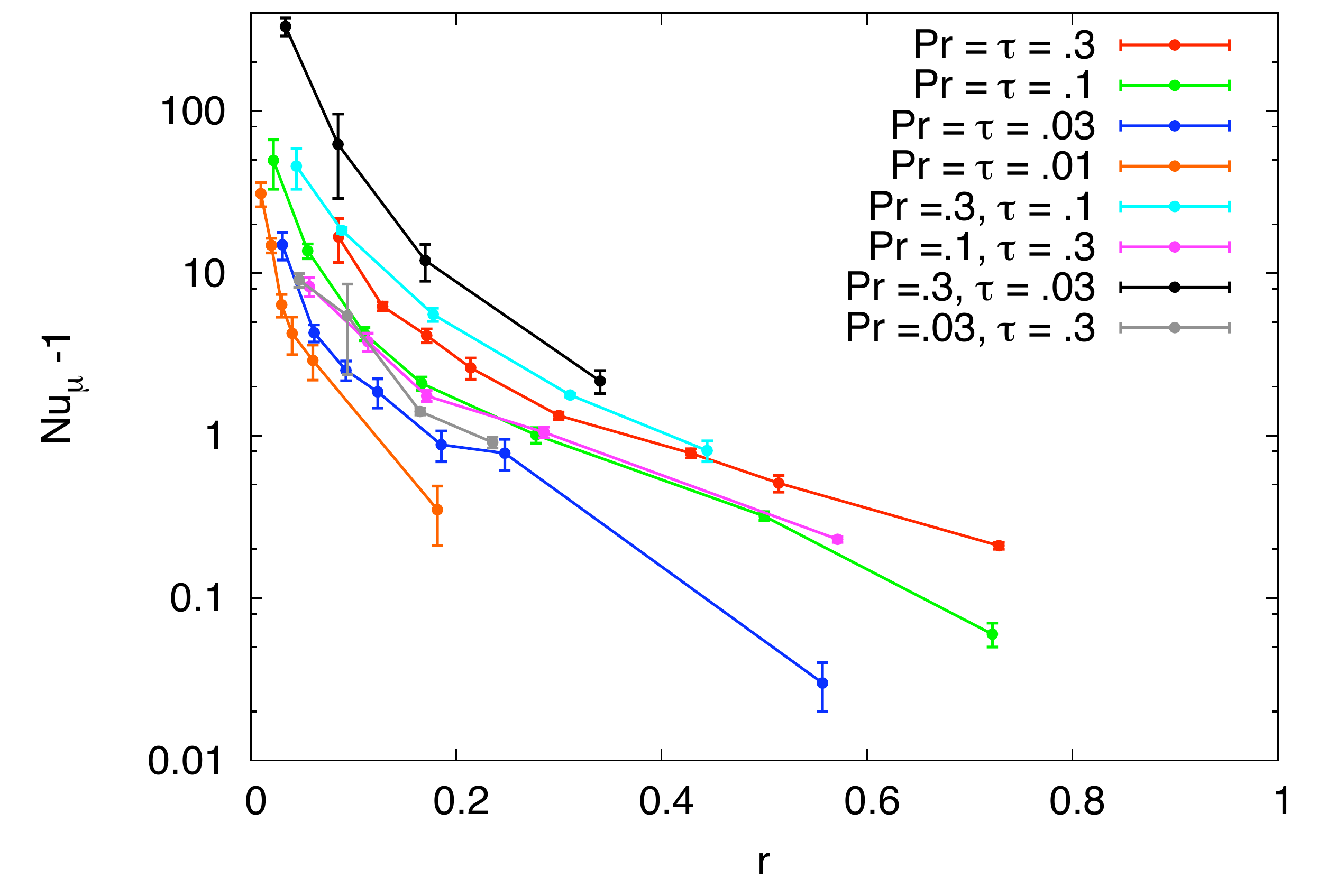}\caption{\small (a) The thermal Nusselt number proxy as a function of the reduced stability parameter $r$ as defined in the main text.  (b) The compositional Nusselt number as a function of the reduced stability parameter $r$. }
\label{fig:nusr}
\end{figure}

Figures \ref{fig:nusr}a and \ref{fig:nusr}b show ${\rm Nu}_T-1 $ and ${\rm Nu}_\mu -1 $ respectively. 
Each curve represents one parameter pair
$({\rm Pr},\tau)$, and is plotted against the stratification parameter $r$, where 
\begin{equation}
r = \frac{R_0^{-1} - 1}{R_c^{-1} - 1 }
\mbox{  . }
\label{eq:littler}
\end{equation}
This quantity is introduced, following \citet{traxler2011b}, to
re-map the instability range into the interval $[0,1]$, with $r=0$
corresponding to Ledoux criterion ($r<0$ being unstable to overturning
convection) and $r=1$ corresponding to the marginal stability limit
($r>1$ being fully stable). This new variable eases the comparison between the various
datasets, and can be interpreted as a rescaled bifurcation
parameter which measures the distance to stability/overturning
instability. 

Figure \ref{fig:nusr} is reminiscent of a similar figure obtained by
\citet{traxler2011b} in the fingering regime. The thermal Nusselt number proxy
is of the order of a few tens, and the
compositional Nusselt number is of the order of a few
hundreds for systems which are nearly 
Ledoux-unstable. Both rapidly drop to one close to the marginal 
stability limit ($r\rightarrow 1$, $R_0^{-1} \rightarrow R_{\rm
  c}^{-1})$. Since a real Nusselt number can also be viewed
(in the Boussinesq limit) as the ratio of the effective diffusivity (turbulent + microscopic) to the microscopic
diffusivity, with
\begin{equation}
D_{\rm eff} = {\rm Nu}_\mu \kappa_{\mu} \mbox{,}
\end{equation}
our results show that turbulent compositional transport can be significant for more unstable systems. 
An equivalent interpretation for heat transport is more delicate, since ${\rm Nu}_T$ can only be viewed 
as a Nusselt number when $\grad_{\rm ad} = 0$. 

\begin{figure}[h!]
\includegraphics[width=0.5\textwidth]{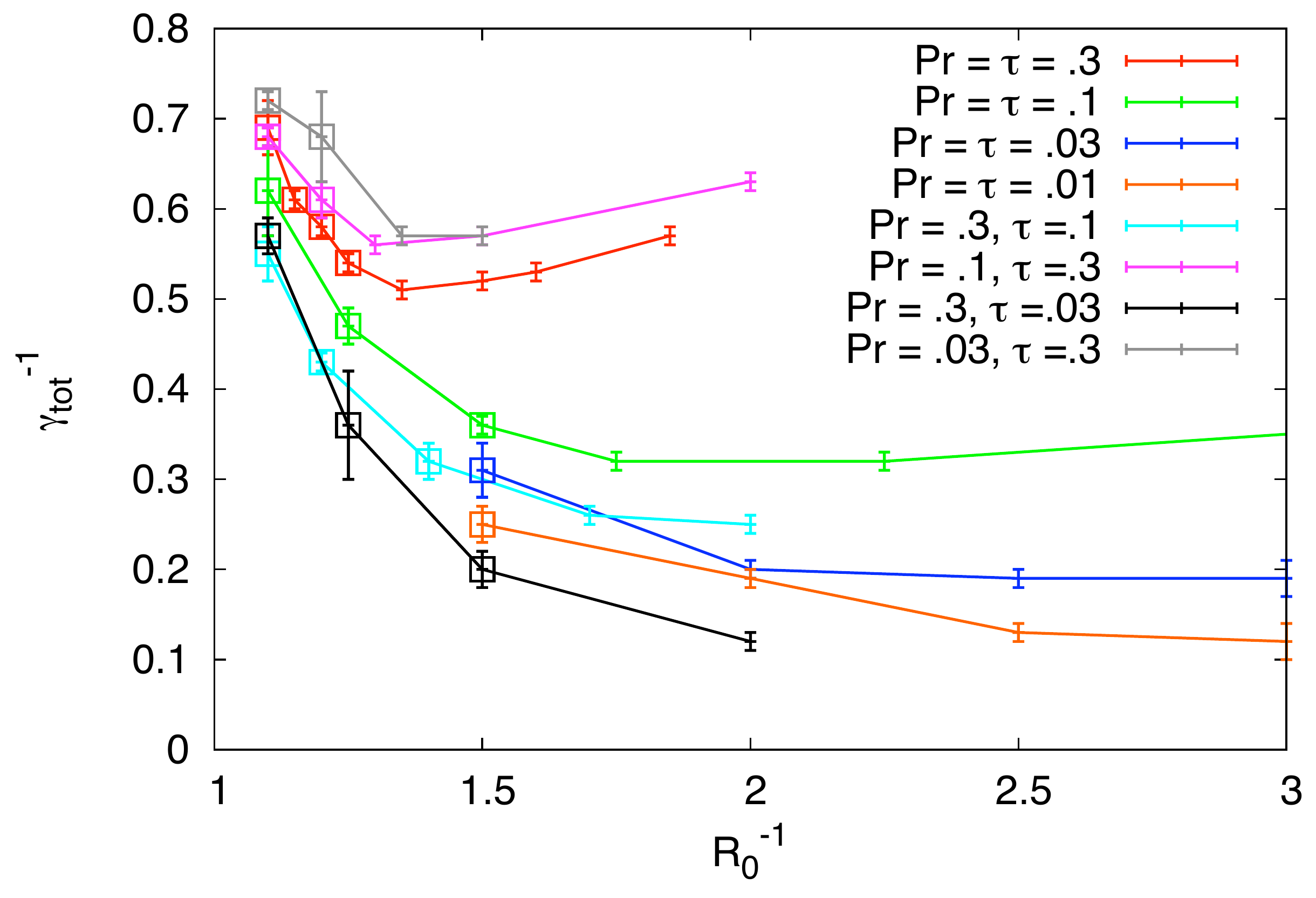}  \includegraphics[width=0.5\textwidth]{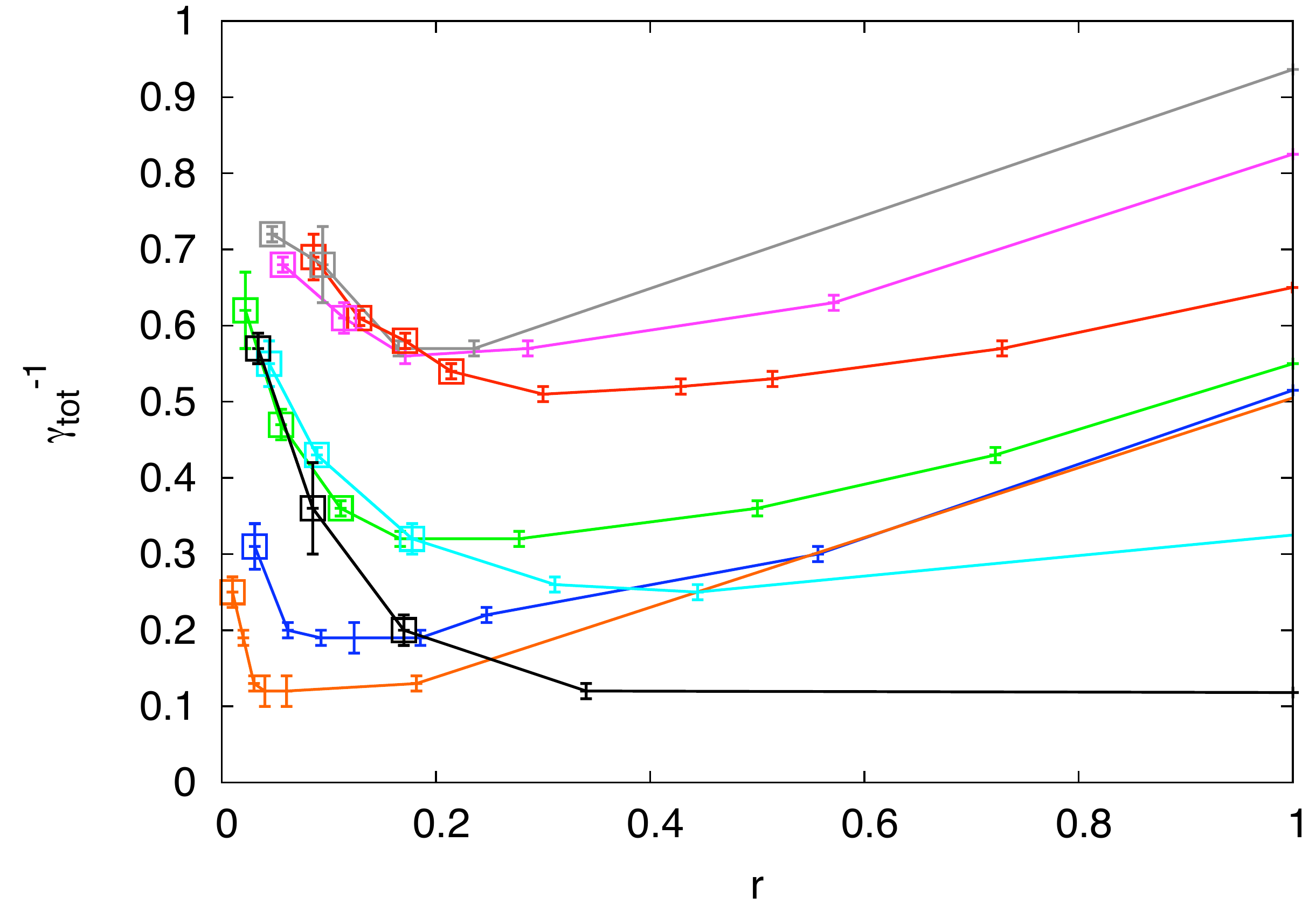}  
\caption{\small (a, left) The flux ratio $\gamma_{\rm tot}^{-1} $ obtained using the averaging methods discussed in Appendix B, for various values of Pr and $\tau$, as a
function of $R_0^{-1}$. Only the interval $R_0^{-1} \in [1,3]$ is
shown to emphasize the region of decreasing $\gamma_{\rm tot}^{-1} $. The larger symbols indicate which 
runs eventually lead to layer formation. (b, right) The same results
plotted against the instability parameter $r$ as defined in the main
text. The value of $\gamma_{\rm tot}^{-1} $ for $r=1$ is the ratio of
the diffusive fluxes, $\gamma_{\rm tot}^{-1}  (r=1) = R_0^{-1} \tau$.}
\label{fig:gamma}
\end{figure}

Figures \ref{fig:gamma}a and \ref{fig:gamma}b  show the flux
ratio $\gamma_{\rm tot}^{-1} $ measured in the simulations, as a function of
$R_0^{-1}$ and as a function of $r$ respectively. Both figures
reveal many interesting features. We find
that for all $({\rm Pr},\tau)$ explored, there exists a region where $\gamma_{\rm tot}^{-1} $ decreases with
$R_0^{-1}$ (equivalently, with $r$), hence, where layer formation is
possible according to the $\gamma-$instability theory. 
We can therefore immediately compare our theoretical expectations
with the actual outcome of the simulations:
Figure \ref{fig:gamma}a and \ref{fig:gamma}b 
show runs which lead to layer formation as larger symbols. 
For larger values of Pr and $\tau$ (i.e. Pr, $\tau$ equal to 0.3
or 0.1), we confirm that layers indeed
form whenever $\gamma_{\rm tot}^{-1} $ is a decreasing function of
$R_0^{-1}$, hence validating the adequacy of Radko's
criterion. For lower Pr and $\tau$, computational constraints limit our ability to 
validate Radko's theory as systematically as in the higher Pr and $\tau$ case. 
Indeed, the layering mode growth rate depends on the derivative of
$\gamma_{\rm tot}^{-1}(R_0^{-1})$ (see Section \ref{sec:newtheory}), 
so the emergence of layers can be delayed significantly in runs with values of
$R_0^{-1}$ close to the minimum of the curve (see for example Figure
\ref{fig:lotsofruns} for $\Pr=\tau=0.1$, $R_0^{-1} = 1.5$). Since simulations at lower 
values of Pr and $\tau$ require considerable spatial resolution, we were not always able to integrate 
them for enough time to see the emergence of layers. We {\it have} seen them for very low values of $R_0^{-1}$ 
where the $\gamma-$curve decreases most rapidly, and expect that they should appear 
for slightly higher values of $R_0^{-1}$ as well.  

The fact that layering is possible in diffusive convection at
low $\Pr$ and $\tau$ is in stark
contrast with results from the fingering regime \citep{traxler2011b}, where $\gamma_{\rm tot}$
always seems to increase with $R_0$ in the same limit. This rather
remarkable difference in behavior is actually fairly easy to 
understand. Indeed, let us first look at the behavior of the
$\gamma-$curve close to marginal stability. In the corresponding runs,
turbulent transport becomes negligible (see Figure \ref{fig:nusr}),
so the flux ratio is dominated by diffusive
transport. Mathematically speaking,  
\begin{eqnarray}
{\rm Nu}_T , {\rm Nu}_\mu \sim 1 \Rightarrow  \gamma_{\rm tot}^{-1}  =
\frac{\tau}{R_0} \frac {{\rm Nu}_\mu}{{\rm Nu}_T} \sim \tau R_0^{-1} = 
  \frac {\tau(1-\tau)}{\Pr + \tau} r + \tau \mbox{   ,}
\end{eqnarray}
which explains the observed oblique asymptote up to the limiting
diffusive value $\tau R_c^{-1}$ at $r=1$ (see Figure
\ref{fig:gamma}b). 
In the diffusive regime considered here, $\tau R_c^{-1} $ is always smaller than one. However a similar argument
applies in the fingering regime and yields $\gamma_{\rm tot} = \tau^{-1} R^{\rm
  fingering}_{\rm c} =  \tau^{-2} \gg 1$. This limit ``pulls up'' the end of the $\gamma-$curve to very
large values, effectively preventing the existence of a region where $ \gamma_{\rm tot}^{-1}$ decreases with $r$.



\section{Theoretical predictions for the flux ratio}
\label{sec:gammapred}

The numerical simulations we have been able to perform sample 
parameter space reasonably comprehensively for $\Pr$ and $\tau$ between 0.01
and 0.3, in particular for values of $R_0^{-1}$ close to unity. 
We found that for all parameter pairs $(\Pr,\tau)$ studied, there
exists a interval $R_0^{-1} \in [1, R_{\rm L}^{-1}]$ where the 
function $\gamma_{\rm tot}^{-1} (R_0^{-1})$ decreases, and that
spontaneous layer formation indeed occurs in that region as expected
from the $\gamma-$instability theory. 
However, in order to create a model for diffusive (semi-) convection that can be used
practically and efficiently in a planetary or stellar evolution code, 
it would be preferable to have an analytical or semi-analytical theory 
for the position $R_{\rm L}^{-1}$ of the minimum of the $\gamma-$curve, 
rather than having to rely on interpolations or
extrapolations of the available dataset presented in Tables 6 and 7.  
In this section, we propose such a model. 

\subsection{Theoretical model for $\gamma_{\rm tot}^{-1}$}
\label{sec:gammatheor}

While we are looking for a model of the flux ratio $\gamma_{\rm tot}^{-1} $, it is interesting to note that a method for estimating the {\it turbulent} flux ratio
\begin{equation}
\gamma^{-1}_{\rm turb} = \frac{\langle \tilde w \tilde \mu \rangle}{\langle \tilde w \tilde T \rangle}
\end{equation}
from linear theory was first proposed by \citet{schmitt1979fgm} in the context of fingering convection. 
Schmitt's theory adequately captures the shape of the curve $\gamma_{\rm turb}(r)$
measured from laboratory \citep{schmitt1979fgm} and numerical experiments
\citep{traxler2011b}, and in particular its dependence on Pr and
$\tau$, although the exact value of $\gamma_{\rm turb}$ for a given
value of $r$ could be off by 20-40\%. As such, it should be considered
as a qualitatively accurate indicator of scalings and trends, but is quantitatively reliable only within factors of ``a few''. 

Schmitt's method can straightforwardly be applied to diffusive convection,
as derived in Appendix A3. The resulting expression for
$\gamma^{-1}_{\rm turb} $ is given by equation (\ref{eq:gammam1turb}),
and depends on the growth rate and wavenumber of the most rapidly
growing mode according to linear theory at the selected parameters $R_0^{-1}$, Pr and $\tau$. 
The latter can be found numerically quite easily by solving
simultaneously a cubic and a quadratic equation. 
If $\gamma^{-1}_{\rm turb}$ is known, then 
\begin{equation}
\gamma_{\rm tot}^{-1} = \frac { \tau R_0^{-1}  + \gamma^{-1}_{\rm
    turb} \overline{\langle \tilde w \tilde T \rangle }}{ 1 +
  \overline{\langle \tilde{w} \tilde{T} \rangle} } = \frac { \tau
  R_0^{-1}  + \gamma^{-1}_{\rm turb} ({\rm Nu}_T-1)}{ 1 + ({\rm Nu}_T-1) } \mbox{   ,}
\label{eq:gammatotlong}
\end{equation}
where we have used equation (\ref{eq:ntgammadef}) to express the turbulent
heat flux in terms of ${\rm Nu}_T$. All that remains to do is to
create a model for ${\rm Nu}_T$ as a function of the system parameters
$R_0^{-1}$, Pr and $\tau$. 

We now return to the results of the numerical simulations presented in
Section \ref{sec:gammares}. We find that we can satisfactorily fit the
behavior of ${\rm Nu}_T-1$ for large values of $r$ provided ${\rm Nu}_T-1 \propto (1-r)$. Close to overturning instability, on the other
hand, we find that a  first satisfactory fit to the data has ${\rm
  Nu_T}-1 \propto (1-\tau)/(R_0^{-1} -1)$. This functional dependence
is not unexpected,  since $R_0^{-1} -1$ is the non-dimensional
background density gradient, and since diffusive convection
relies on $\tau \neq 1$ to operate, and is
much more efficient the smaller the value of $\tau$. Combining this
fit with the large $R_0^{-1}$ limit suggests a functional form with
${\rm Nu}_T-1 \propto (1-r)(1-\tau)/(R_0^{-1} -1)$. One can 
fit the proportionality constant for runs with Pr = $\tau$, and obtain
a rather good match to the data. However, the resulting expression is
less satisfactory for Pr $\neq \tau$. 
Further investigation reveals that an even better fit can be obtained with 
\begin{equation}
\langle \tilde{w}\tilde{T} \rangle = {\rm Nu}_T - 1 = (0.75 \pm 0.05) \left(\frac{\Pr }{\tau}\right)^{ 0.25 \pm 0.15}
\frac{1-\tau}{R_0^{-1}-1} (1-r) \mbox{   .}
\label{eq:NuTmod}
\end{equation}
The large uncertainty on the power index of the term $(\Pr/\tau)$ comes from the uncertainty on the
measurements themselves, compounded with the short range of Pr/$\tau$
values available. However, its exact value does not matter
much for the $\gamma_{\rm tot}^{-1}$ predictions. 
\begin{figure}[h!]
\centerline{\includegraphics[width=0.6\textwidth]{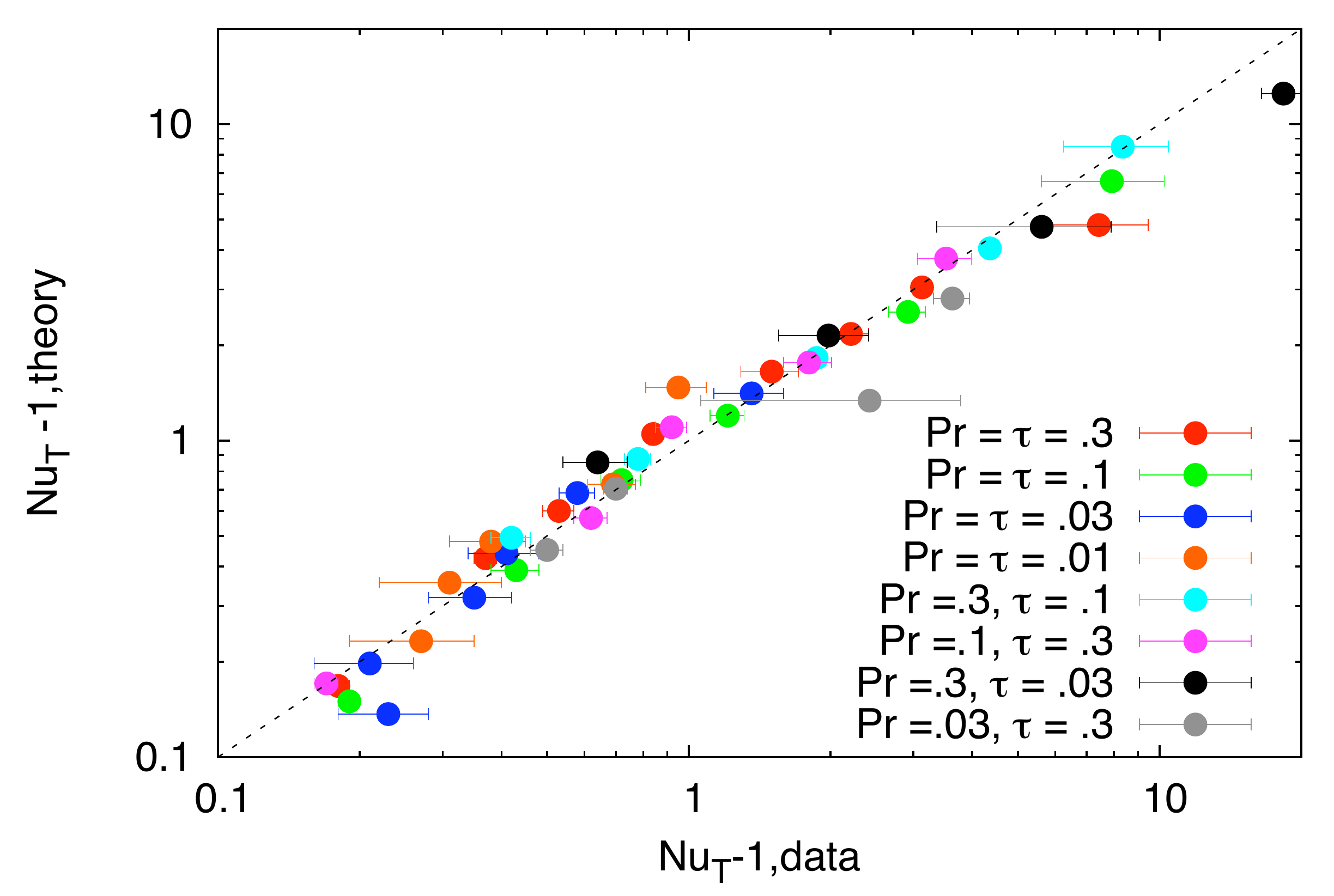}  }
\caption{\small Comparison between expression (\ref{eq:NuTmod}) and
  the data presented in Tables 6 and 7. }
\label{fig:NuTpred}
\end{figure}

Figure
\ref{fig:NuTpred} compares our empirical fit for ${\rm Nu_T}-1$ given
by equation (\ref{eq:NuTmod}) to the actual data. This fit is
satisfactory for our current purposes, although we recognize that a
better theoretically-motivated one should be sought in the future if
we wish to improve on the model further. 

\begin{figure}[h!]
\centerline{\includegraphics[width=0.6\textwidth]{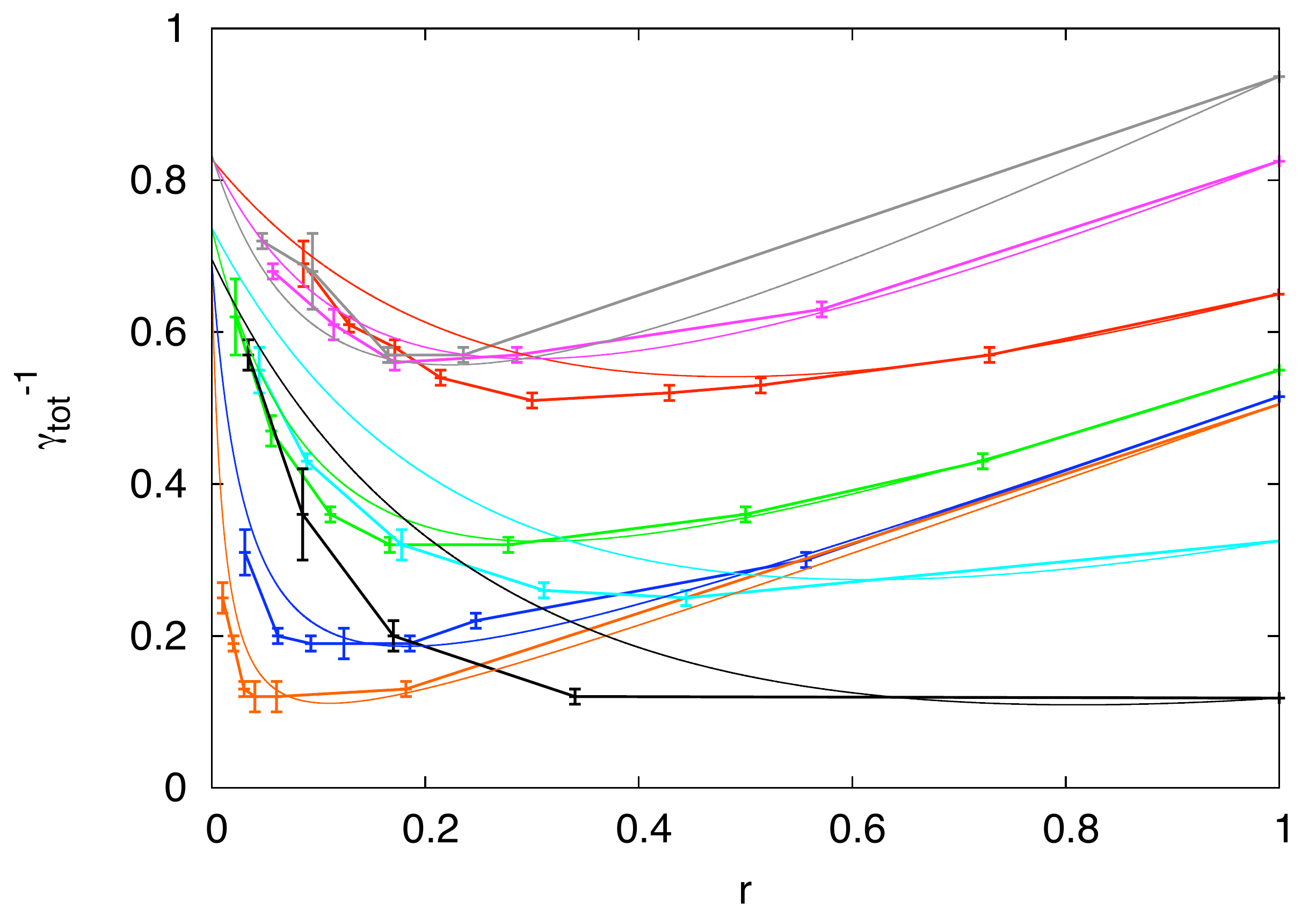}  }
\caption{\small Comparison between the model prediction for
  $\gamma_{\rm tot}^{-1} (r)$ and the data presented in Tables 6 and 7 and Figure
  \ref{fig:gamma}. The same color scheme is used in this figure to represent the various parameter pairs $(\Pr,\tau)$ 
as in Figures 6, 7, and 9.}
\label{fig:gammatheor}
\end{figure}

Using (\ref{eq:gammam1turb}), (\ref{eq:gammatotlong}) and (\ref{eq:NuTmod}) we can now estimate $\gamma_{\rm tot}^{-1}$
semi-analytically. Figure \ref{fig:gammatheor} compares our predictions with the data presented in Figure \ref{fig:gamma}b. 
As expected from the limitations of Schmitt's method, and uncertainties
in our fit for the turbulent heat flux, the model does not match the data perfectly. 
Generally speaking, we find that the predicted value of $r$ at the
minimum is somewhat overestimated by the model, by 20-40\%. The slope
of the $\gamma-$curve is thus also affected. These discrepancies are
larger and/or more apparent for runs with larger values of the Prandtl
number, for which the position of the minimum occurs for larger values
of $r$.  However, it is nevertheless rather remarkable to see how well
our model accounts for the shape of the $\gamma-$curve, and in
particular the variation of the position of the minimum 
with Pr and $\tau$.  The value of $\gamma_{\rm tot}^{-1} $ at the minimum is also
robustly predicted by the model. 

\subsection{Model trends}

\begin{figure}[h!]
\centerline{\includegraphics[width=0.8\textwidth]{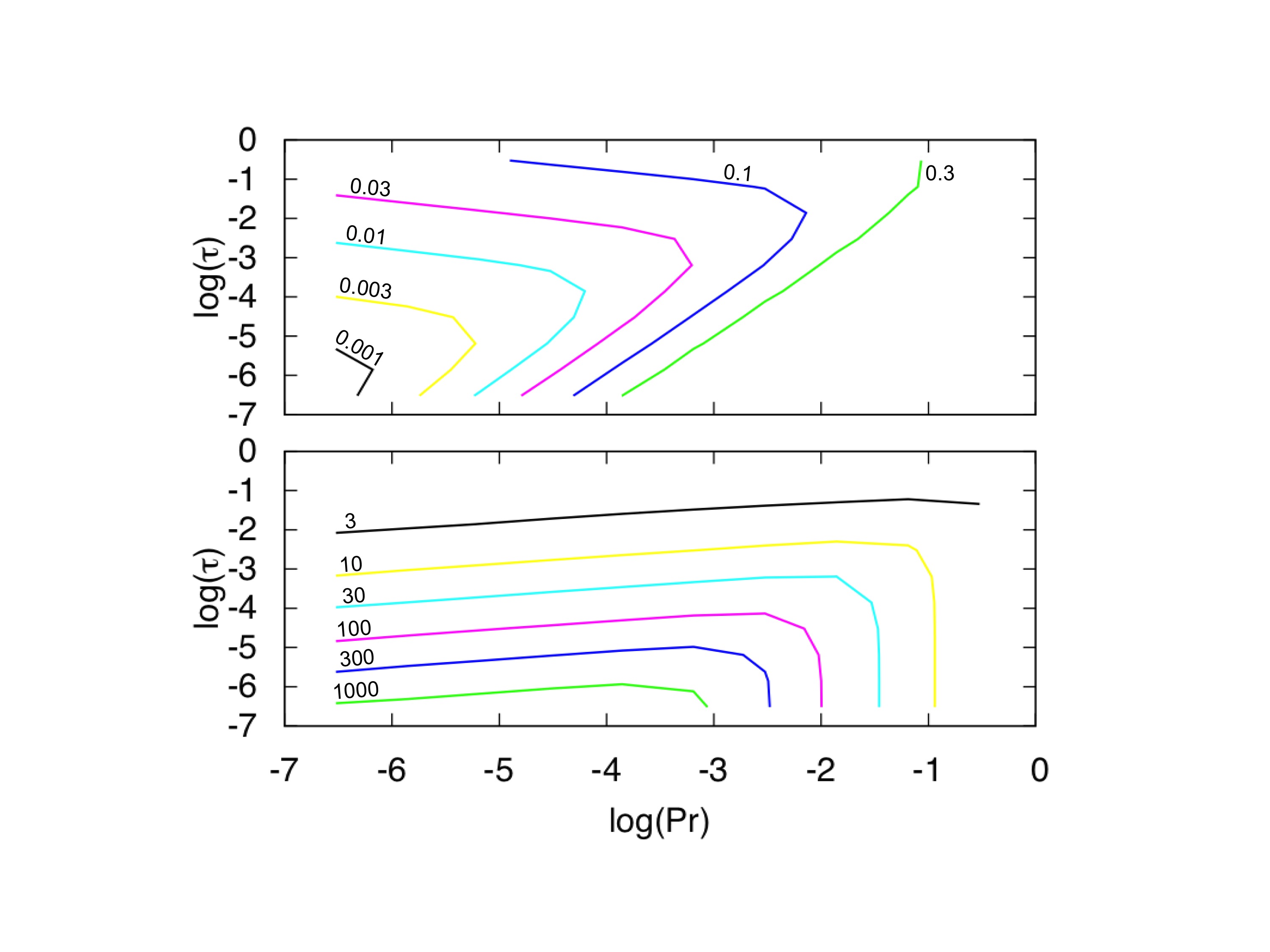} } 
\caption{\small Contour plots of the position of the minimum of the $\gamma-$curve as a function of Pr and $\tau$, written as 
  $R_L^{-1}$ (bottom) and $r_L = (R_L^{-1}-1)/(R_c^{-1}-1)$ (top). Note that both $r_L$ and $R_L^{-1}$ calculated directly from the model, as shown here, are likely to overestimate the true position of the minimum by about 20\%-40\%. } 
\label{fig:rmin}
\end{figure}

Using the model described in the previous section, we can now 
estimate the value of the inverse density ratio $R_L^{-1}$ for which 
$\gamma_{\rm tot}^{-1}$ is minimal, for a range of parameter
values beyond those for which we were able to run numerical simulations.
The results are presented in Figure \ref{fig:rmin}, for Pr and
$\tau $ varying from $10^{-7}$ to 1, and show contour plots of
$R_L^{-1}$ (bottom) and of the corresponding $r_L = (R_L^{-1}-1)/(R_c^{-1}-1)$ (top). 
Note that both $r_L$ and $R_L^{-1}$ estimated directly from the model, as shown here, are likely to overestimate the true position of the minimum by about 20\%-40\% (see previous section). 

Overall we find that the relative fraction of the total instability range
unstable to layering decreases as Pr and $\tau$
decrease (i.e. the value of $r$ at the 
minimum of the $\gamma-$curve decreases). However, since the
instability range itself increases as Pr and $\tau$ tend to zero, the value $R_L^{-1}$
below which layers can spontaneously emerge actually increases significantly. We
find that it is of the order of a few for planetary values of the
diffusivities, and of the order of a few hundreds to a thousand for
the stellar parameter regime. The layering instability, and its implications on 
increasing the heat and compositional transport properties of diffusive convection, is thus
likely to play an important role in stellar and planetary
astrophysics. 
 
\section{Comparison of the layer growth rates with theory}
\label{sec:layers}

\citet{rosenblum2011} already showed that Radko's $\gamma-$instability
theory, when applied to the case of diffusive convection,
correctly accounts for the growth rate of the emergent staircase in
their simulations at $\Pr = \tau = 0.3$ with $R_0^{-1} = 1.2$. In this
section, we check that this is still true at lower values of $\Pr$ and
$\tau$, and compare two methods for estimating the mode growth rate:
one based on the experimentally-determined functions $\gamma_{\rm
  tot}^{-1}(R_0^{-1})$ and ${\rm Nu}_T(R_0^{-1})$ listed in Tables 6 and 7, and one based on
the model functions proposed in Section \ref{sec:gammapred}. 

As described in Section \ref{sec:newtheory} the growth rate $\Lambda$
of a layering-mode with vertical wave-number $k$ is the solution of
the quadratic (\ref{eq:newgammaeq}). Estimating $\Lambda$
thus requires first estimating ${\rm Nu}_0 = {\rm Nu_T}(R_0^{-1})$ and
$\gamma^{-1}_0 = \gamma_{\rm tot}^{-1}(R_0^{-1})$ respectively, as well as the derivative terms $A_1$
and $A_2$ defined in equations (\ref{eq:A1def}) and (\ref{eq:A2def})
respectively. 
This can either be done using the actual experimental data, or using
our new semi-analytical theory (see Section \ref{sec:gammapred}). 
When using the experimental data, $A_1$ and $A_2$ are calculated using either one-sided or two-sided
derivatives, depending on the data points available. When using the semi-analytical
model functions, $A_1$ and $A_2$ are always calculated using two points on either sides of the selected value of $R_0^{-1}$. 

To illustrate the process, we compare the $\gamma-$instability theory with the data from the 
$\Pr = \tau =0.03$, $R_0^{-1} = 1.5$ run. Table \ref{tab:layergrowth1} shows the results
of our estimates for the layering mode growth rate $\Lambda$ using the two different methods. The
experimentally-derived results are expected to be more accurate since they
do not rely on any modeling. Reassuringly, however, we find that the model-derived growth rate is within
20\% of the experimentally derived one. It is interesting to note that
while the model-estimate for $A_2$ seems to be off by an order unity, this discrepancy
does not affect the growth rate estimate much. This remark is valid
for many of the cases studied (although in many other cases $A_2$ is
well-predicted by the theory).  

\begin{table}[!h]
\begin{center}
\begin{tabular}{ccc}
\hline
 &  Experiment (M1) &  Model (M2) \\
\hline
\hline
${\rm Nu}_0$  &  2.36 & 2.41  \\
$\gamma_0^{-1}$  &  0.31  & 0.38  \\
$A_1$  & 0.33  & 0.49 \\
$A_2$  & 2.34  & 4.36 \\
$\Lambda/k^2 $   & 0.31  & 0.36 \\
\hline
\end{tabular}
\caption{Layering mode growth rate using ${\rm Nu}_0$,
  $\gamma_0^{-1}$, $A_1$ and $A_2$ from the experimental data (M1) and
  from the model presented in Section \ref{sec:gammapred} (M2)
  respectively, for the run with $\Pr = \tau = 0.03$, $R_0^{-1} =
  1.5$. Note that since $\Lambda$ is proportional to $k^2$, we list
  the proportionality constant $\Lambda/k^2$ for more generality. }
\label{tab:layergrowth1}
\end{center}
\end{table}

\begin{figure}[h!]
\centerline{\includegraphics[width=0.6\textwidth]{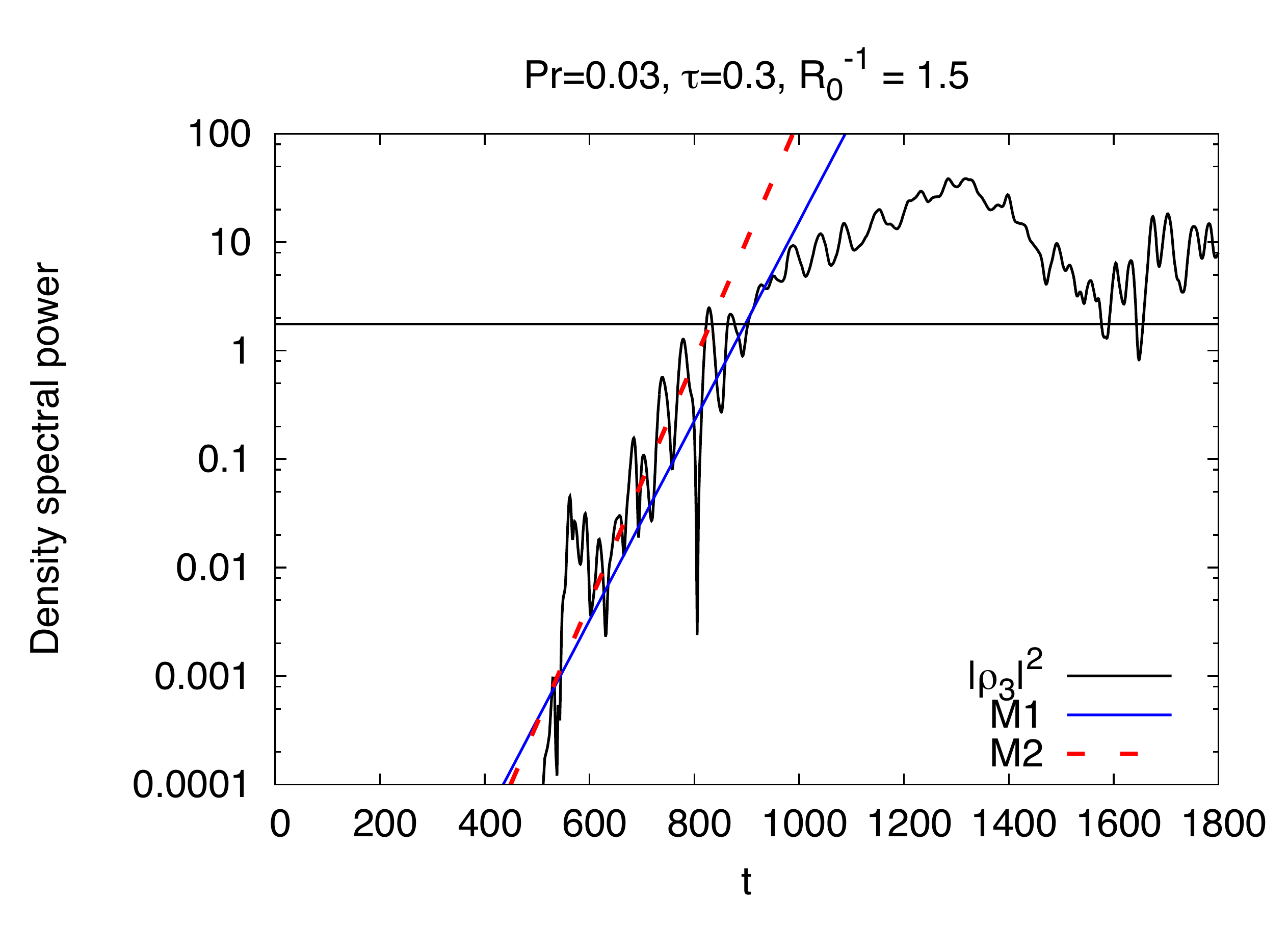} } 
\caption{\small Comparison between the model prediction for the
  layering mode growth rate and the actual data, for $\Pr = \tau =
  0.03$, $R_0^{-1} = 1.5$. The emergent mode observed has $n=3$ steps so
  $k = 3 (2\pi/L_z)$. The case using the growth rate derived from the
  experimental data is shown as M1, while the one using the growth rate
  derived from the model for the thermal Nusselt number proxy and flux ratio
  proposed in Section \ref{sec:gammapred} is shown as M2. Also shown
  is the critical amplitude for onset of overturning convection, as a
  horizontal line. The mode growth notably changes upon reaching this
  amplitude, and quickly saturates after that. } 
\label{fig:Layer1}
\end{figure}

Figure \ref{fig:Layer1} compares our estimates for $\Lambda$ with the actual mode growth
observed in the simulations. As shown by \citet{stellmach2011} and \citet{rosenblum2011}, a
convenient way of extracting the amplitude of the layering mode is to
look at the Fourier expansion of the density field, and isolate the
mode with zero horizontal wavenumber, and a vertical wavenumber $k =
n (2\pi/L_z)$ where $n$ is the number of steps in the emergent
staircase. Figure \ref{fig:Layer1} shows the square of the norm of that mode
(i.e. its power $|\hat \rho_n|^2$), and compares it with an exponential function
proportional to $e^{2 \Lambda t}$ (the normalization being
arbitrary). Both growth rate estimates correctly account for the
observed mode growth, with the experimentally-derived growth rate
faring somewhat better, as expected. However, it is reassuring to see that
the model proposed in Section \ref{sec:gammapred} works quite well too. 

Once the mode's amplitude grows beyond a certain critical value, its
density profile is no longer monotonously decreasing. When this
happens, localized regions become unstable to overturning convection,
and a fully-formed staircase rapidly appears. The threshold for
overturning instability for a mode with $n$ steps was calculated by
\citet{rosenblum2011} to be, in terms of its density spectral power, 
\begin{equation}
|\hat \rho_n|^2_{\rm conv} = \left( \frac{1-R_0^{-1}}{2n} \right)^{2} \mbox{    .   }
\end{equation}
Figure \ref{fig:Layer1} clearly shows that the mode growth rapidly stops 
after its amplitude crosses that threshold. 

Applying the same method to all runs which eventually result in
layer formation, we find that the layer growth rate predicted by
the solution of equation (\ref{eq:newgammaeq}) always correctly accounts for the observed mode
growth in the simulations. Furthermore, while the growth
rate predicted from experimentally-derived values of ${\rm Nu}_0$, $\gamma_0^{-1}$,
$A_1$ and $A_2$ is always better than the model-derived ones, the
latter are nevertheless satisfactory estimates too, and are always
within 10-30\% of the correct value. These results complete the validation of Radko's theory, as well as our 
model estimates for $\gamma_{\rm tot}^{-1}$ and ${\rm Nu}_T$.

\section{Conclusion and prospects}
\label{sec:ccl}

The ultimate goal of this series of papers is to propose a new model
for mixing by diffusive convection. In the work presented here,
we ran and analyzed a very extensive suite of numerical simulations of
the process in a wide range of parameter space. We have 
shown that in the astrophysically-relevant low Prandtl number ($\Pr = \nu/\kappa_T$), low
diffusivity ratio ($\tau = \kappa_\mu/\kappa_T$) regime, diffusive convection can take one of two forms
depending on the local inverse density ratio $R_0^{-1} =
\grad_\mu/(\grad - \grad_{\rm ad})$: moderately efficient layered
convection at lower
$R_0^{-1}$, or inefficient wave-dominated ``oscillatory'' convection
for higher $R_0^{-1}$ (see Figure \ref{fig:layerregime}). We have
confirmed through numerical and analytical work that a spontaneous transition into layered convection occurs, under predictable circumstances,
through a linear mean-field instability of the initial state of
oscillatory convection. This instability is called the
$\gamma-$instability. It was originally suggested in the oceanographic 
context of fingering convection by \citet{radko2003mlf} and later applied to double-diffusive
convection in astrophysics by \citet{rosenblum2011}. 

Since transport in layered convection is much more efficient than in
the non-layered case, a crucial element of any new model of
diffusive convection will be the availability of a practical
criterion for determining, for given $R_0^{-1}$, $\Pr$ and $\tau$, whether a system is expected to transition
into layers or not, and on what timescale. The original
$\gamma-$instability theory provides such a criterion, but the latter 
can only be used in practice provided experimental
measurements of the turbulent fluxes in homogeneous diffusive
convection, for the same parameters, are available. We provide such
measurements here for the planetary parameter regime, but similar
results are unlikely to ever be available for the much more extreme
stellar parameter regime. 

Based on these considerations, we then proposed a new empirically
motivated model for the turbulent fluxes, which enable us to derive a
completely parameter-free semi-analytical criterion to determine, for
{\it any} given fluid in the astrophysical regime ($\Pr, \tau \ll 1$) and any given stratification ($R_0^{-1}$),
(a) whether a system is expected to transition into layers or not, and (b) on
what timescales the layers are expected to emerge. Our model was found
to fit the available numerical data very well, and can therefore be
used very reliably within the same region of parameter space (e.g. the
planetary parameter regime, for which $\Pr, \tau \sim
10^{-2}-10^{-1}$). We further propose 
that it should also be used in regions of parameter space for which
fully resolved simulations are not available, namely in the stellar parameter regime ($\Pr, \tau \sim
10^{-7}-10^{-5}$). Based on this model, we find that layered convection
is theoretically expected in stellar interiors for a fairly wide range of parameter
space, with $R_0^{-1}$ between 1 and about 1000. 

Our results answer, at least approximately, the first of the three questions we initially raised: (1) Under which conditions do layers form? (2) What is the transport rate in layered convection and (3) What is the transport rate in non-layered ``oscillatory'' convection. In subsequent papers in this series we will continue our investigation by answering questions (2) and (3).





\acknowledgments

We thank Nic Brummell, Jonathan Fortney, and Douglas Gough for fruitful discussions. G.M., P.G. and T.W. were supported by funding from the NSF (NSF-0807672), and benefited from the hospitality of the ISIMA program during the summer of 2011. A.T. P.G. and T.W. were funded by the NSF (NSF-0933759). 
Part of the computations were performed on the UCSC Pleiades supercomputer,
purchased with an NSF-MRI grant. Others used computer resources at the National Energy Research Scientific Computing Center (NERSC), which is supported by the Office of Science of the US Department of Energy under contract DE-AC03-76SF00098. Figure 1 was rendered using ViSiT. P.G. thanks LLBL Hank Childs for his excellent support of the software. 

\appendix 

\section{Appendix: Linear stability of semi-convection, asymptotic solutions for low Pr
  and $\tau$, fastest-growing modes}

The system of equations (\ref{eq:goveqs}) can be linearized and solved
for the fastest growing modes of diffusive convection. These linear
solutions can then be studied further to obtain asymptotic scalings at
very low Pr and $\tau$, and to derive predictions for the turbulent buoyancy
flux ratio (see Section \ref{sec:gammapred}).  

\subsection{Linearized equations for the fastest-growing modes}

We first linearize (\ref{eq:goveqs})
around $\tilde{T} = \tilde{\mu} = 0$ and ${\bf \tilde{u}}=0$, 
assuming all perturbations are normal modes of the form $\tilde{q} =
\hat{q} e^{ilx + imy + ikz + \lambda t}$ where $q \in \left\{ T, \mu,
    \bf{u} \right\}$. Hatted quantities 
are now the amplitudes of the perturbations, while $l$ and $m$ are the horizontal wave-numbers, $k$ is the vertical
 one, and $\lambda$ is the growth rate. The latter are all
 non-dimensional. 
\\ 

We are interested in the fastest-growing modes only, which
can be shown to have $k=0$ as in the fingering case
\citep{radko2003mlf,traxler2011b}. They correspond to purely vertical
fluid motions. They are rotationally invariant around the vertical
direction, so without loss of generality we can align 
the horizontal wavenumber with the $x$-axis choosing $m=0$. 
After some simplifications, the resulting system of equations for the mode amplitudes are
\begin{eqnarray}
\lambda \hat T - \hat w = - l^2 \hat T \mbox{   ,} \nonumber \\
\lambda \hat \mu - R_0^{-1} \hat w = -\tau l^2 \hat \mu \mbox{   ,} \nonumber \\
\lambda \hat w = -  \Pr l^2 \hat w + \Pr(\hat T-\hat \mu) \mbox{   .}
\label{eq:linearpert}
\end{eqnarray}
It has a non-trivial solution provided the growth
rate $\lambda$ satisfies the following cubic equation:
\begin{equation}
\left(\frac{\lambda}{\Pr}+ l^2 \right) (\lambda + l^2) (\lambda + \tau l^2) - (\lambda + \tau l^2)+R_0^{-1} (\lambda + l^2) = 0 ~.
\label{eq:cubic}
\end{equation}

In the regime of interest, this cubic has 
one negative real root and two complex conjugate roots
\citep{baines1969}. It can easily be shown that 
the complex conjugate roots ($\lambda = \lambda_R + i \lambda_I$ with $\lambda_I \ne 0$) 
satisfy
\begin{equation}
\lambda_I^2 = 3\lambda_R^2 + 2 l^2  \lambda_R \left( \tau+ \Pr + 1
\right)  + l^4 \left( \tau  + \Pr\tau + \Pr
  \right) + \Pr (R_0^{-1} -1 )  \mbox{   ,}
\label{eq:lambdaI}
\end{equation}
and that $\lambda_R$ satisfies the cubic 
\begin{eqnarray}
8 \lambda_R^3 + 8 l^2 \lambda_R^2 ( \tau+ \Pr + 1 ) 
\nonumber \\ + 2 \lambda_R \left[    l^4 \left( \tau  + \Pr\tau + \Pr
  +\left( \tau+ \Pr + 1
\right)^2   \right) +  \Pr (R_0^{-1} -1 )  \right] \nonumber \\ 
+  l^6 (\tau + \Pr)(\tau+1) ( \Pr + 1)  +   l^2 \Pr (R_0^{-1} \left( \tau+ \Pr \right) - \left(
     \Pr + 1 \right) ) 
=0 \mbox{   .}
\label{eq:lmax1}
\end{eqnarray}
The latter has a positive solution if and only if $R_0^{-1} \in \left[
  1, R_c^{-1}\right]$ where $R_c^{-1} = \frac{\Pr+1}{\Pr + \tau}$. 

The fastest growing modes are determined by fixing Pr, $\tau$ and $R_0^{-1}$ within the instability range, 
and finding the value of $l$ for which $\lambda_R$ is 
maximum by solving (\ref{eq:lmax1}) in conjunction with $\frac {d
  \lambda_R }{d l^2} = 0$, or in other words
\begin{eqnarray}
8 \lambda_R^2 ( \tau+ \Pr + 1 ) + 4 \lambda_R l^2 \left( \tau  + \Pr\tau + \Pr
  +\left( \tau+ \Pr + 1
\right)^2   \right)  \nonumber \\ 
+  3l^4 (\tau + \Pr)(\tau+1) ( \Pr + 1)  + \Pr (R_0^{-1} \left( \tau+ \Pr \right) - \left(
     \Pr + 1 \right) ) 
=0 \mbox{   .}
\label{eq:lmax2}
\end{eqnarray}

In what follows, we study the behavior of the solutions as a function of
 the reduced stratification parameter $r$, defined in
 (\ref{eq:littler}). Note that, with this new variable, we have 
\begin{equation}
R_0^{-1} \left( \tau+ \Pr \right) - \left(
     \Pr + 1 \right) = (r-1)(1-\tau) \mbox{   .}
\end{equation}

\subsection{Asymptotic solutions at low Pr and $\tau$}

In general, one needs to solve (\ref{eq:lmax1}) and (\ref{eq:lmax2})
numerically to find the fastest growing modes for given $ {\rm Pr},\tau$ and $R_0^{-1}$. 
Here, however, we are interested in
deriving asymptotic solutions for low Pr and low $\tau$, since this is the
parameter regime relevant for planetary and stellar interiors. In particular, we want to study how the growth rate and the wavenumber of
 the fastest-growing modes scale with these governing parameters.

\begin{figure}
\includegraphics[width=0.5\textwidth]{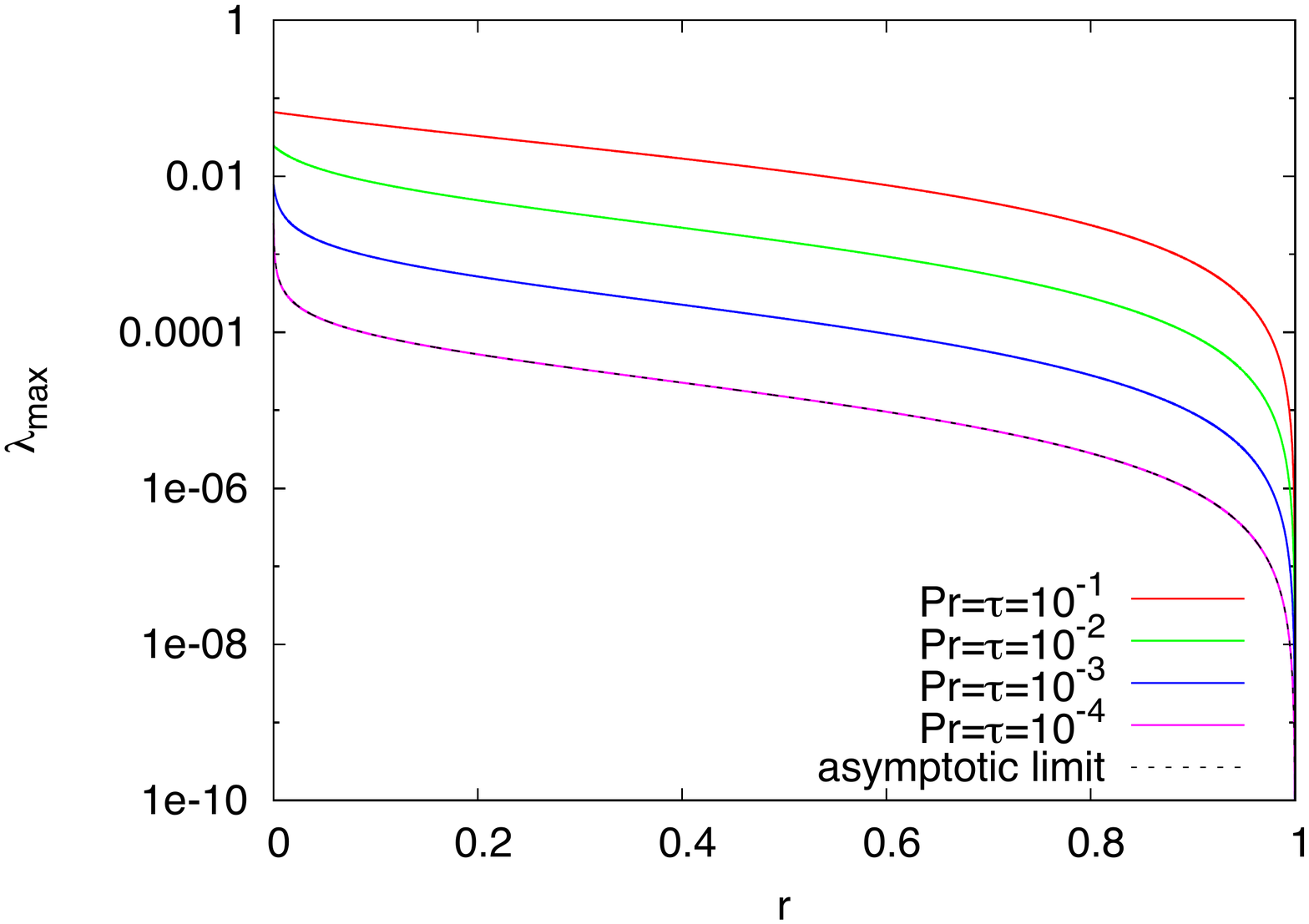}  \includegraphics[width=0.5\textwidth]{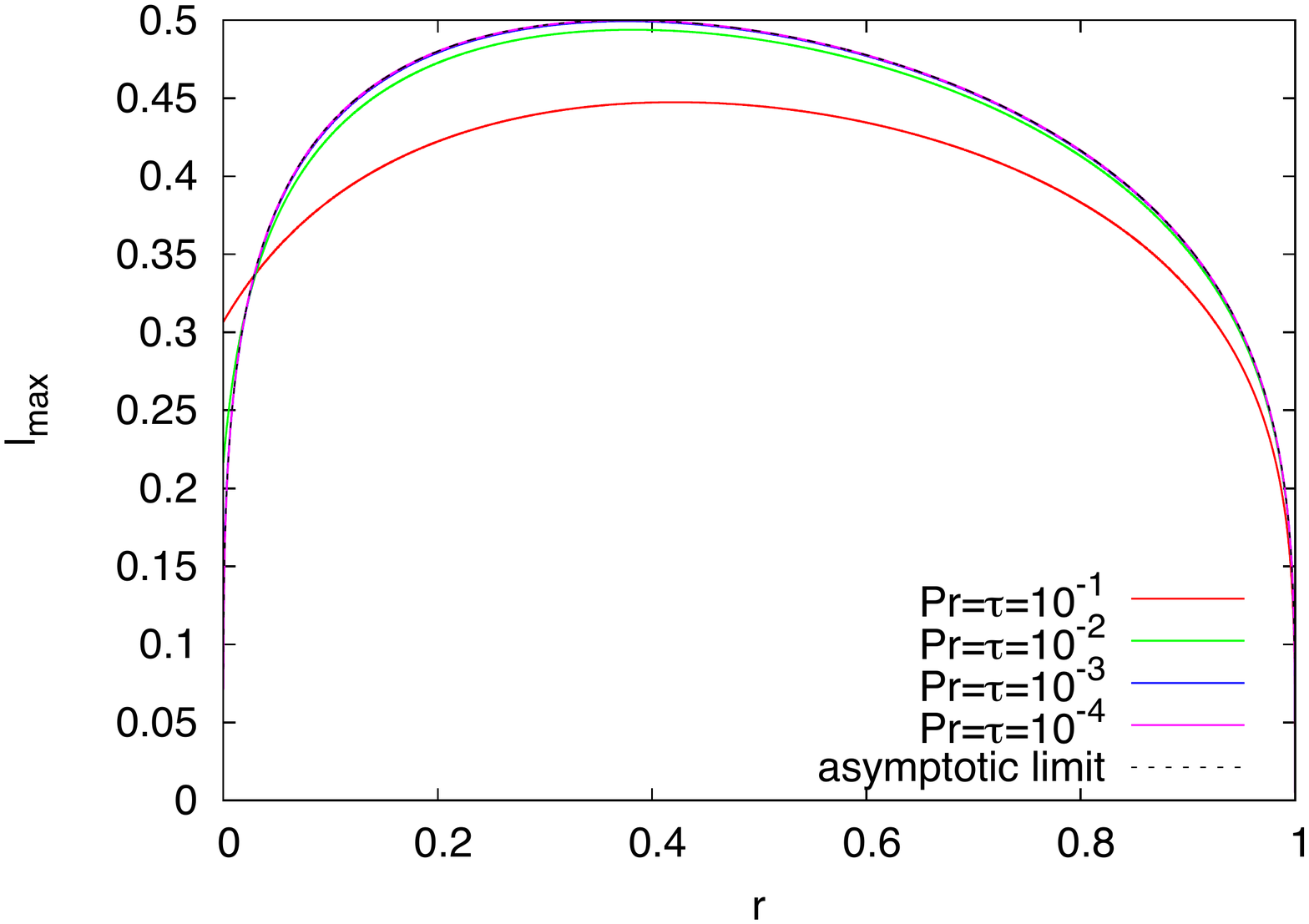}
\caption{\small Plots for the non-dimensional growth rate
  $\lambda_{\rm max}$  (left) and
  the non-dimensional horizontal
  wavenumber $l_{\rm max}$ (right) of the fastest-growing mode, for
  Pr$=\tau$. 
Left: We see that $\lambda_{\rm max}$ scales with Pr. 
The dotted line shows the asymptotic solution of (\ref{eq:asympsol}),
multiplied by $\Pr$. 
Right: The horizontal wavenumber rapidly becomes independent of
Pr. The fastest-growing wavelength is $2\pi/l_{\rm max}$, so of the
order of 13-20$d$ at planetary parameter regimes. The dotted line shows the asymptotic solution of (\ref{eq:asympsol}).}
\label{fig:Prscaling}
\end{figure}

The solutions to equations (\ref{eq:lmax1}) and (\ref{eq:lmax2}) can easily be found numerically, and
the results are shown in Figure \ref{fig:Prscaling} for decreasing
Pr (here with Pr$=\tau$). We see that the real part of the fastest growing mode's
non-dimensional growth rate, $\lambda_{\rm max}(r)$, 
appears to be proportional to Pr, while the corresponding horizontal
wavenumber, $l_{\rm max}(r)$, becomes independent of Pr as Pr decreases. 

This behavior suggests a new rescaling of the governing equations to
capture the asymptotic regime ($\Pr, \tau \rightarrow 0$) 
of the instability: $\lambda_{\rm max} = \Pr \hat \lambda $ where
$\hat\lambda \sim O(1)$, and $l_{\rm max} = \hat l$ where $\hat l \sim
O(1)$. 
We also define $\phi = \tau/\Pr$, and assume that $\phi$ is order
unity. Using (\ref{eq:littler}) and keeping only the lowest terms in
$\Pr$, equations (\ref{eq:lmax1}) and (\ref{eq:lmax2}) reduce to the
simple universal\footnote{Note that this asymptotic limit is not uniformly valid for
  $r \rightarrow 0$.}  system:
\begin{eqnarray}
 4 \hat l^2 \hat{\lambda} + 3\hat l^4 (\phi + 1) + (r-1) = 0 \nonumber
 \\
2\left(\hat l^4 + \frac{r}{\phi + 1}\right)\hat{\lambda} +\hat l^6
(\phi + 1) + \hat l^2(r-1) = 0
\label{eq:asympsol}
\end{eqnarray}
This system still needs to be solved numerically for $\hat \lambda$
and $\hat l$, but only once for each value of $\phi$ and $r$. 
Figure \ref{fig:Prscaling} compares the solution of
(\ref{eq:asympsol}) with the ones obtained by direct numerical
solution of (\ref{eq:lmax1}) and (\ref{eq:lmax2}) for $\Pr = 10^{-4}$
and $\phi = 1$, and confirms our numerical and semi-analytical results. 

This linear asymptotic analysis helps us estimate the growth rate of
this kind of double-diffusive instability for a broad range of parameters 
and determine the size of the basic unstable structures we are
interested in. Dimensionally speaking, our results imply that the true
lengthscale of the instability should always be of the order of a few
$d$, where $d$ was defined in (\ref{eq:nondims}), and the growth rate
of the instability should be of the order of $\Pr \kappa_T/d^2 =
\sqrt{\Pr} N$ where $N$ is the thermal buoyancy (Br\"unt-V\"ais\"al\"a)
frequency. This information is useful for two 
reasons: first, to quantify the expected lengthscales or timescales in the real systems (i.e. stellar and
planetary interiors), and secondly, to get some insight into the
correct domain size and timestep to use in the numerical simulations. 




\subsection{Semi-analytical prediction for the turbulent buoyancy flux ratio}

Let us consider the {\it turbulent} flux ratio
\begin{equation}
\gamma^{-1}_{\rm turb} = \frac{\langle \tilde{w}\tilde{\mu} \rangle}{\langle\tilde{w}\tilde{T}  \rangle}
\end{equation}
where $\langle \cdot \rangle$ denotes a spatial average over the entire computational domain. 
\citet{schmitt1979fgm} showed that it is possible to estimate this quantity
for {\it fingering} convection using the velocity field, temperature and
  chemical composition perturbations corresponding to the linearly
  fastest-growing mode of instability. Since the
  unknown amplitude of the perturbations in this turbulent ratio cancels
  out, the remaining expression only depends on the known
  shape and growth rate of the perturbations. Here, we apply the same
  technique to estimate the turbulent flux ratio in diffusive convection. 

From the system of equations (\ref{eq:linearpert}), we see that the
amplitudes of the vertical velocity, temperature and compositional
perturbations of a given mode are related via
\begin{eqnarray}
\hat T = \frac{\hat w}{\lambda+l^2} \mbox{   ,}  \nonumber \\
\hat \mu = \frac{R_0^{-1} \hat w}{\lambda+\tau l^2} \mbox{   .} 
\label{eq:hattmuw}
\end{eqnarray}
In order to calculate the fluxes, we must remember that, in the
process of the linear analysis, the various fields 
$\tilde{w}$, $\tilde{T}$ and $\tilde{\mu}$ were defined as complex variables, 
e.g. from $\tilde{q} = \hat{q} e^{ilx + imy + ikz + \lambda t}$, under the implicit understanding that only their real parts are physically meaningful. Hence, in practice, 
\begin{equation}
\gamma^{-1}_{\rm turb} = \frac{\langle \Re(\tilde{w})\Re(\tilde{\mu}) \rangle}{\langle \Re(\tilde{w})\Re(\tilde{T}) \rangle}\mbox{   .} 
\end{equation}
Without loss of generality, $\hat w$ can be selected to be real so that 
\begin{equation}
\Re(\tilde{w}) = \hat w e^{\lambda_R t } \cos(lx + \lambda_I t) \mbox{   .} 
\end{equation}
Then, using (\ref{eq:hattmuw}), we find that 
\begin{eqnarray}
\Re(\tilde{T}) && = \frac{\hat w e^{\lambda_R t}}{(\lambda_R + l^2 )^2  + \lambda_I^2 } \left[ \cos(lx+\lambda_I t) (\lambda_R + l^2) + \sin(lx+\lambda_It) \lambda_I \right] \mbox{   ,} \nonumber \\
\Re(\tilde{\mu}) && = \frac{R_0^{-1} \hat w e^{\lambda_R t}}{(\lambda_R + \tau l^2 )^2  + \lambda_I^2 } \left[ \cos(lx+\lambda_I t) (\lambda_R + \tau l^2) + \sin(lx+\lambda_It) \lambda_I \right]\mbox{   .} 
\end{eqnarray}

Finally, forming the turbulent flux ratio and integrating the relevant
quantities over the computational domain and over short timescales
(i.e. over at least one oscillation period of the basic instability), we get $\gamma^{-1}_{\rm
  turb}(l,\lambda)$ for a given mode with wavenumber $l$ and growth rate
$\lambda = \lambda_R + i \lambda_I$ as 
\begin{equation}
\gamma^{-1}_{\rm turb}(l,\lambda) = R_0^{-1}\frac{(\lambda_R + l^2 )^2  + \lambda_I^2 }{(\lambda_R + \tau l^2 )^2  + \lambda_I^2 }\frac{\lambda_R + \tau l^2   }{\lambda_R + l^2   }\mbox{   .} 
\label{eq:gammam1turb}
\end{equation}
where $\lambda_R$ and $\lambda_I$ are related via (\ref{eq:lambdaI}). 
Applying this formula to the most rapidly growing mode, with
wavenumber $l_{\rm max}$ and growth rate $\lambda_{\rm max}$
calculated in Appendix A1, yields the required estimate for the inverse turbulent flux ratio in our simulations. 


\section{Appendix B: Extraction of mean fluxes and results}

\subsection{Protocol for extracting mean fluxes and measuring $\gamma_{\rm tot}^{-1}$ from the simulations}

In what follows, we describe our protocol for measuring the mean turbulent fluxes in the homogeneous phase of diffusive convection (prior to the emergence of large-scale structures). This involves first creating a systematic method to identify the start and end times $[t_{\rm start}, t_{\rm end}]$  of this phase and then estimating the fluxes and related errorbars.

\subsubsection{Selection of $t_{\rm start}$.}

\begin{figure}[h]
\centerline{\includegraphics[width=0.5\textwidth]{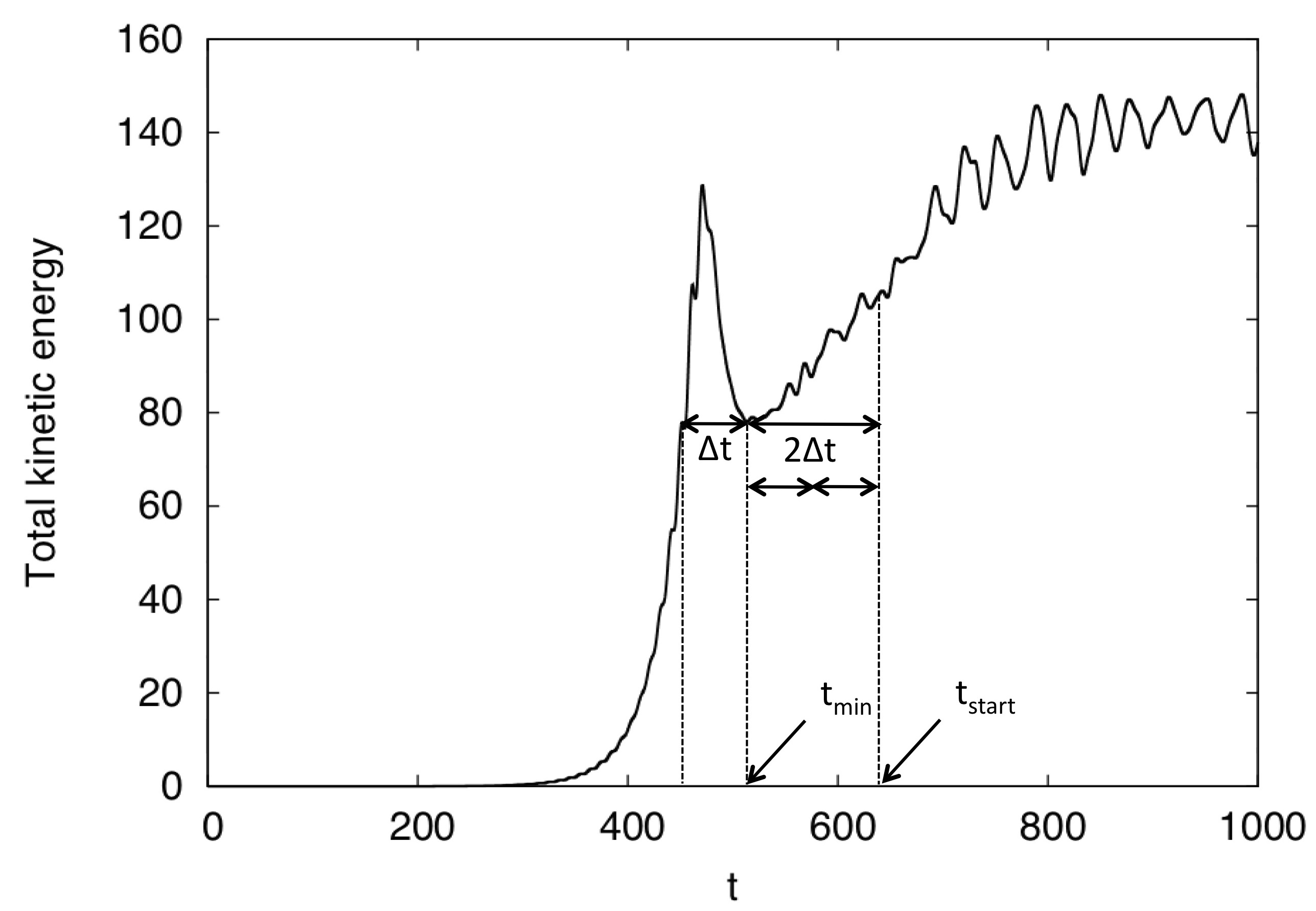}}
         \caption{\small Illustration of the method used to select
           $t_{\rm start}$, applied to the simulation with $\Pr = \tau
           = 0.1$, $R_0^{-1} = 1.75$. We first find the time $t_{\rm min}$ when
           the total kinetic energy (post saturation), has its first local minimum. We then define the width of the peak $\Delta t$
           as the time interval elapsed since the last time the total
           kinetic energy had the same value. Finally, we define
           $t_{\rm start} = t_{\rm min} + 2 \Delta t$.     }
\label{fig:tstart}
\end{figure}

As seen in Figure \ref{fig:lotsofruns}, the turbulent flux typically
peaks then drops quite sharply during the saturation of the primary
instability, and then grows more slowly towards its value in the
homogeneous double-diffusively convecting state. As shown in Figure
\ref{fig:tstart}, the same description applies to the behavior of the
total kinetic energy in the system. It thus appears that
the system needs a little bit of time to ``recover'' from the
saturation. In order to extract meaningful averages, we therefore need to
select the start of the averaging process well-past the main
saturation peak. We also need to define $t_{\rm start}$ in a manner
that is meaningful across all simulations. Figure \ref{fig:tstart} illustrates our
process: we define first the ``width'' of the saturation peak $\Delta
t$ as illustrated, and then choose $t_{\rm start}$ accordingly,
about $2\Delta t$ past the peak. While
this choice is arguably somewhat arbitrary, it does satisfy the
requirements listed above. Furthermore, the estimated values of
$\gamma_{\rm tot}^{-1}$ are not particularly sensitive to the choice
of $t_{\rm start}$ as long as it is indeed well-past the saturation
peak. 

\subsubsection{Selection of $t_{\rm end}$ in the non-layered case}

As noted earlier, by contrast with \citet{rosenblum2011} we find that
even in the non-layered case the system does not necessarily remain in
a state of homogeneous, small-scale diffusive convection but
sometimes becomes dominated by larger-scale coherent gravity
waves\footnote{\citet{rosenblum2011} did not notice the
  emergence of the waves in their simulations, although a more careful
  re-analysis of their results shows that they were
  indeed present in some of the higher $R_0^{-1}$ runs.}. While the precise reason for the emergence and synchronization of these waves remains to
be determined, their associated dynamics lead to a rather different
type of transport than in more homogeneous diffusive
convection. For this reason, we must identify when the waves first ``take
over'' and restrict our measurements of the turbulent fluxes prior to that time. 

\begin{figure}[h]
\centerline{\includegraphics[width=0.6\textwidth]{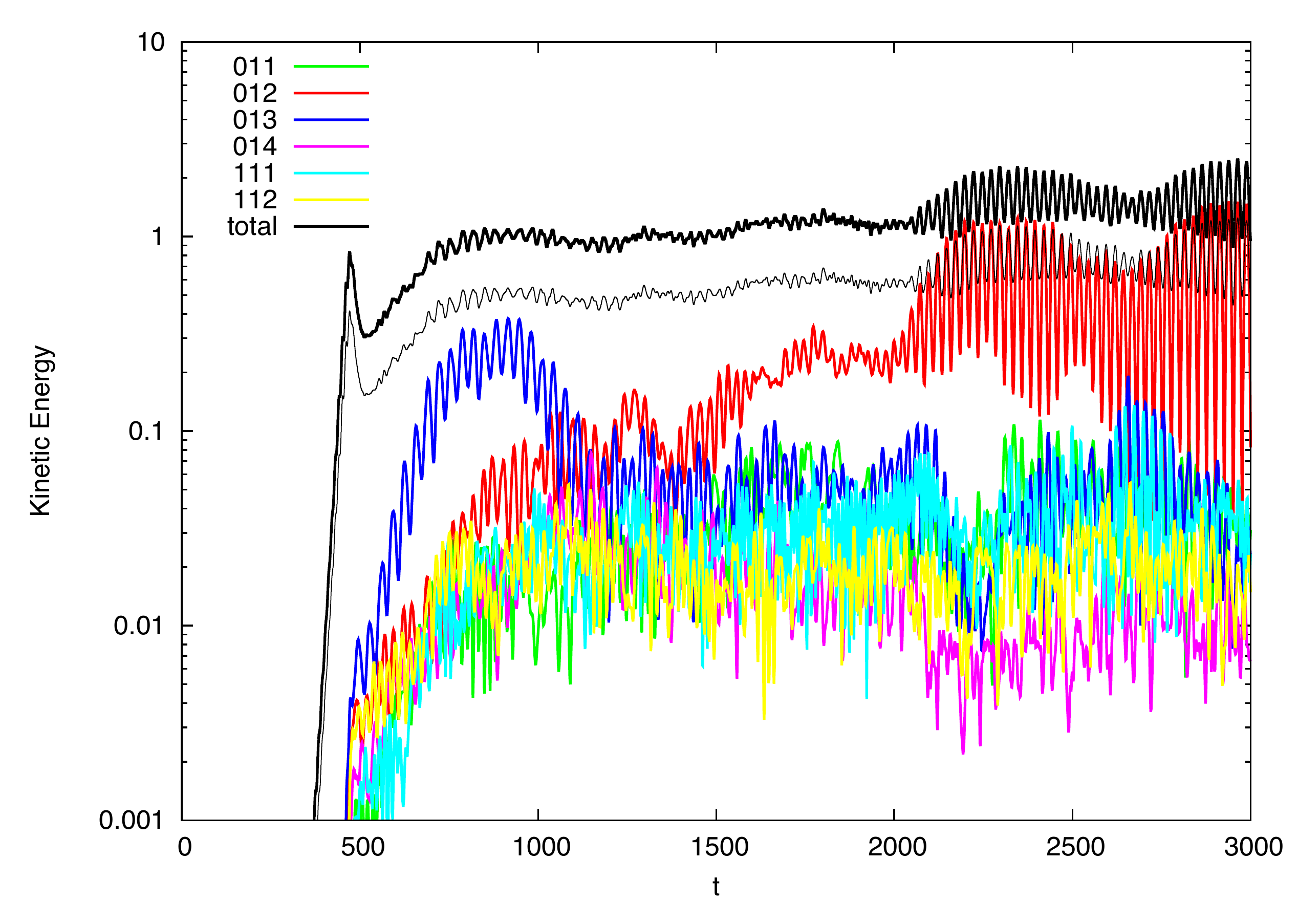}}
         \caption{\small Analysis of the simulation with $\Pr = \tau =
           0.1$, $R_0^{-1} = 1.75$. This plot shows the temporal evolution of the total kinetic
           energy in the system (thick black line), half that quantity
           (thin black line), as well as the total kinetic energy in
           the six highest-amplitude gravity wave mode  families (see
           main text for definition and notation). Around $t=2000$ the
         kinetic energy in the 012 mode family reaches 1/2 the total
         kinetic energy in
       the system, at which point it clearly begins to dominate
       the system's transport rates.  }
\label{fig:waves}
\end{figure}

Shown in Figure \ref{fig:waves} is the total kinetic energy in the
simulation, as well as the total kinetic energy in the six
highest-amplitude families of gravity wave modes. By ``families'', we
imply the following. A single gravity-wave mode, in this
triply-periodic simulation, can be identified with the Fourier
mode proportional to $\exp(ik_x x + ik_y y + i
k_z z)$, where $(k_x,k_y,k_z)$ is the mode wave-vector. A ``family'' of modes is defined as the ensemble of all the modes with the
same geometry given the symmetries of the system, i.e. the same values
of $|k_z|$ and the same values of $|k_h| = \sqrt{k_x^2 + k_y^2}$. In
what follows, we classify the modes for simplicity of notation based on their
periodicity: the single mode $\{0,2,-1\}$ for example corresponds to
one with $k_x = 0$, $k_y = 2(2\pi/L_y)$, $k_z = -(2\pi/L_z)$. The
family of modes 021 then corresponds to an ensemble of 8 modes: $\{0,2,1\}$, 
$\{0,2,-1\}$, $\{0,-2,1\}$, $\{0,-2,-1\}$, $\{2,0,1\}$, 
$\{2,0-1\}$, $\{-2,0,1\}$ and finally $\{-2,0,-1\}$. Finally, the
total kinetic energy in the mode family is just the sum of that of the
individual modes. 

Figure \ref{fig:waves} shows that the evolution of the total kinetic
energy of the system is very similar to that of the turbulent fluxes
for the same simulation (see Figure
\ref{fig:lotsofruns}): an extended, apparently quasi-steady turbulent
state between $t \sim 600$ and $t \sim 2000$, followed by a
wave-dominated phase. It also reveals that the family
of modes which dominates the system beyond $t=2000$ is the 012
family, and that the strong oscillatory signal in the total kinetic energy (and
the turbulent heat flux) appears when the total kinetic energy in that
single family exceeds half the total kinetic energy of the system
(shown as the thin black line). 

We used a similar method to analyze every single simulation among the
ones presented in Tables 1 and 2, comparing the total kinetic energy to that
of various families of modes, and found that a robust (albeit empirical)
criterion for determining the time $t_{\rm gw}$ when a system becomes dominated by gravity
waves is simply that the total kinetic energy in any given family of modes
exceeds half the total kinetic energy in the system. 
In the non-layered case, we therefore take the ``end-point'' $t_{\rm
  end}$ of the temporal average to be $t_{\rm end} =  t_{\rm gw}$.  In
some cases with $R_0^{-1}$ close to marginal stability, it can happen
that the start and end times thus selected have $t_{\rm start} \ge
t_{\rm end}$. When this is the case, we discard the simulation (for
the purpose of estimating the turbulent fluxes and their ratio). 

\subsubsection{Selection of $t_{\rm end}$ in the layered case}

Applying the method described in the previous section we find that, in
runs which eventually show the emergence of a staircase, gravity-wave modes never dominate the
system. However, since we are interested here in the process which leads to layer formation,
extracting the flux ratio in the layered case is only
meaningful {\it prior} to the formation of the first layers. So,
whenever layers appear in the simulations, we set $t_{\rm end}$ to be
the time where the first set of layers appears. 


\subsubsection{Averaging method and error estimates}

Once the relevant time interval has been determined, we need to
measure the mean turbulent fluxes, construct ${\rm Nu}_T$, ${\rm Nu}_\mu$ and 
$\gamma_{\rm tot}^{-1} $ and estimate our experimental error. 
For this purpose, we use a ``4-intervals'' method:  we first divide
the integration domain previously defined into four sub-intervals, and
calculate the mean fluxes and therefore ${\rm Nu}_T$, ${\rm Nu}_\mu$
and $\gamma_{\rm tot}^{-1} $ in each one of them according to (\ref{eq:ntgammadef}) and (\ref{eq:numudef}). 
The final adopted value of ${\rm Nu}_T$, ${\rm Nu}_\mu$
and $\gamma_{\rm tot}^{-1} $ respectively is then the average of the four
computed values, while the error is their standard deviation. The
reason for using this method is clarified in the examples below.

Let us first illustrate our procedure on the data from the simulation
shown in Figure \ref{fig:kinds}a, i.e. for the run that leads to layer
formation (with $\Pr = \tau=0.03$, $R_0^{-1} = 1.5$). We first estimate the start- and 
end-times of the homogeneous phase to be $t_{\rm start} =775 $ and $t_{\rm end} =
1200$. The mean ${\rm Nu}_T$, ${\rm Nu}_\mu$ and $\gamma_{\rm tot}^{-1} $ in each sub-intervals are given in Table
\ref{tab:gamma1.5}, as well as their final values and corresponding
errorbars. These results illustrate the reason for using such a
method to estimate the measurement ``error'' rather than a simple
average over a single interval: the mean ${\rm Nu}_T$ and ${\rm Nu}_\mu$ increase
steadily from one sub-interval to the other, showing that the system
is not actually in a statistically quasi-steady state (as assumed by
the $\gamma-$instability theory). However, this clearly does not
prevent the layering modes from growing anyway, and as shown in Section
\ref{sec:layers}, the theory still adequately accounts for their growth
rate despite the non-stationarity of the homogeneous
phase. As such, we have to do the best with the data we have, and
report on the values of ${\rm Nu}_T$, ${\rm Nu}_\mu$ and $\gamma_{\rm tot}^{-1} $ accordingly,
albeit with large errorbars which account for the slow temporal evolution
of the system from saturation to the emergence of the staircase.  

\begin{table}[!h]
\begin{center}
\begin{tabular}{cccccc}
\hline
&$ t_{\rm start} $ & $ t_{\rm end} $ & ${\rm Nu}_T$  & $ {\rm Nu}_\mu $ & $ \gamma_{\rm tot}^{-1}  $ \\
\hline
\hline
Interval 1&$ 775 $&$ 881.26 $&$ 2.082 $&$ 12.567 $&$ 0.272 $\\
Interval 2&$ 881.26 $&$ 987.5 $&$ 2.256 $&$ 14.531 $&$ 0.290 $\\
Interval 3&$ 987.5 $&$ 1093.75 $&$ 2.384 $&$ 16.519 $&$ 0.309 $\\
Interval 4&$ 1093.77 $&$ 1200 $&$ 2.703 $&$ 20.448 $&$ 0.341 $\\
\hline
Total     &$ 775 $&$ 1200 $&$ 2.36 \pm 0.23 $&$ 15.9 \pm 2.9 $&$ 0.31 \pm 0.03 $\\
\tableline
\end{tabular}
\end{center}
\caption{\small Illustration of our data averaging method for the run
  with $\Pr = \tau = 0.03$, $R_0^{-1} = 1.5$. }
\label{tab:gamma1.5}
\end{table}

Applying this method to the non-layered run shown in
Figure \ref{fig:waves} (with Pr=$\tau=0.1$, $R_0^{-1} = 1.75$) we
find that the start and end of the homogeneous period are $t = 650$
and $t = 2100$, and the corresponding ${\rm Nu}_T$, ${\rm Nu}_\mu$ and $\gamma_{\rm tot}^{-1} $
computed are shown in Table \ref{tab:gamma5}. In this case the run is more
stationary overall, leading to much smaller errorbars. 

\begin{table}[!h]
\begin{center}
\begin{tabular}{cccccc}
\hline
& $ t_{\rm start} $ & $ t_{\rm end} $ & $ {\rm Nu}_T $ & $ {\rm Nu}_\mu $ & $ \gamma_{\rm tot}^{-1}  $ \\
\hline
\hline
Interval 1&$ 650 $&$ 1012.5 $&$ 1.776 $&$ 3.279 $&$ 0.323 $\\
Interval 2&$ 1012.5 $&$ 1375 $&$ 1.639 $&$ 2.830 $&$ 0.302 $\\
Interval 3&$ 1375 $&$ 1737.5 $&$ 1.673 $&$ 2.992 $&$ 0.313 $\\
Interval 4&$ 1737.5 $&$ 2100 $&$ 1.791 $&$ 3.308 $&$ 0.323 $\\
\hline
Total     &$ 650 $&$ 2100 $&$ 1.72 \pm 0.07 $&$ 3.10 \pm 0.20 $&$ 0.32 \pm 0.01 $\\
\tableline
\end{tabular}
\end{center}
\caption{\small Illustration of our data averaging method for the run
  with $\Pr = \tau = 0.1$, $R_0^{-1} = 1.75$.}
\label{tab:gamma5}
\end{table}

\subsection{Summary of the results}.

The results of our analysis are summarized in Tables 6 and 7. 

\begin{table}
\begin{center}
\begin{tabular}{ccccccccc}
\\
\tableline
Pr       &$ \tau  $&$ R_0^{-1} $&$ r   $&$ t_{\rm start} $&$ t_{\rm
  end} $&$ \gamma_{\rm tot}^{-1} $&${\rm Nu}_T $&$ {\rm Nu}_\mu $ \\
\hline
\hline
$   0.3 $&$   0.3 $&$  1.1 $&$  0.09 $&$   356 $&$   506 $&$   0.69 \pm   0.03 $&$   8.43 \pm   2.02 $&$  17.7 \pm   5.1 $ \\
$   0.3 $&$   0.3 $&$  1.15 $&$  0.13 $&$   370 $&$   700 $&$   0.61 \pm   0.01 $&$   4.13 \pm   0.14 $&$   7.26 \pm   0.38 $ \\
$   0.3 $&$   0.3 $&$  1.2 $&$  0.17 $&$   450 $&$   920 $&$   0.58 \pm   0.01 $&$   3.21 \pm   0.20 $&$   5.14 \pm   0.41 $ \\
$   0.3 $&$   0.3 $&$  1.25 $&$  0.21 $&$   450 $&$  2000 $&$   0.54 \pm   0.01 $&$   2.50 \pm   0.21 $&$   3.62 \pm   0.39 $ \\
$   0.3 $&$   0.3 $&$  1.35 $&$  0.30 $&$   550 $&$   660 $&$   0.51 \pm   0.01 $&$   1.84 \pm   0.03 $&$   2.33 \pm   0.07 $ \\
$   0.3 $&$   0.3 $&$  1.5 $&$  0.43 $&$   700 $&$   950 $&$   0.52 \pm   0.01 $&$   1.53 \pm   0.04 $&$   1.78 \pm   0.05 $ \\
$   0.3 $&$   0.3 $&$  1.6 $&$  0.51 $&$   940 $&$  1070 $&$   0.53 \pm   0.01 $&$   1.37 \pm   0.02 $&$   1.51 \pm   0.06 $ \\
$   0.3 $&$   0.3 $&$  1.85 $&$  0.73 $&$  2200 $&$  2830 $&$   0.57 \pm   0.01 $&$   1.18 \pm   0.01 $&$   1.21 \pm   0.01 $ \\
\hline
$   0.1 $&$   0.1 $&$  1.1 $&$  0.02 $&$   400 $&$   480 $&$   0.62 \pm   0.05 $&$   8.92 \pm   2.31 $&$  50.6 \pm  16.7 $ \\
$   0.1 $&$   0.1 $&$  1.25 $&$  0.06 $&$   500 $&$   720 $&$   0.47 \pm   0.02 $&$   3.92 \pm   0.26 $&$  14.7 \pm   1.4 $ \\
$   0.1 $&$   0.1 $&$  1.5 $&$  0.11 $&$   575 $&$  2150 $&$   0.36 \pm   0.01 $&$   2.21 \pm   0.10 $&$   5.24 \pm   0.39 $ \\
$   0.1 $&$   0.1 $&$  1.75 $&$  0.17 $&$   650 $&$  2100 $&$   0.32 \pm   0.01 $&$   1.72 \pm   0.07 $&$   3.10 \pm   0.20 $ \\
$   0.1 $&$   0.1 $&$  2.25 $&$  0.28 $&$   820 $&$  2700 $&$   0.32 \pm   0.01 $&$   1.43 \pm   0.05 $&$   2.01 \pm   0.11 $ \\
$   0.1 $&$   0.1 $&$  3.25 $&$  0.50 $&$  1780 $&$  2050 $&$   0.36 \pm   0.01 $&$   1.19 \pm   0.01 $&$   1.32 \pm   0.02 $ \\
$   0.1 $&$   0.1 $&$  4.25 $&$  0.72 $&$  4300 $&$  4900 $&$   0.43 \pm   0.01 $&$   1.05 \pm   0.01 $&$   1.06 \pm   0.01 $ \\
\hline
$   0.03 $&$   0.03 $&$  1.5 $&$  0.03 $&$   775 $&$  1200 $&$   0.31 \pm   0.03 $&$   2.36 \pm   0.23 $&$  16.0 \pm   2.9 $ \\
$   0.03 $&$   0.03 $&$  2   $&$  0.06 $&$  1000 $&$  1600 $&$   0.20 \pm   0.01 $&$   1.58 \pm   0.05 $&$   5.30 \pm   0.52 $ \\
$   0.03 $&$   0.03 $&$  2.5 $&$  0.09 $&$   620 $&$  1250 $&$   0.19 \pm   0.01 $&$   1.41 \pm   0.07 $&$   3.53 \pm   0.35 $ \\
$   0.03 $&$   0.03 $&$  3   $&$  0.12 $&$  1300 $&$  2215 $&$   0.19 \pm   0.02 $&$   1.35 \pm   0.07 $&$   2.86 \pm   0.38 $ \\
$   0.03 $&$   0.03 $&$  4   $&$  0.19 $&$  1650 $&$  2250 $&$   0.19 \pm   0.01 $&$   1.21 \pm   0.05 $&$   1.88 \pm   0.19 $ \\
$   0.03 $&$   0.03 $&$  5   $&$  0.25 $&$  3400 $&$  4800 $&$   0.22 \pm   0.01 $&$   1.23 \pm   0.05 $&$   1.78 \pm   0.17 $ \\
$   0.03 $&$   0.03 $&$ 10   $&$  0.56 $&$  5700 $&$  6500 $&$   0.30 \pm   0.01 $&$   1.02 \pm   0.01 $&$   1.03 \pm   0.01 $ \\
\hline
$   0.01 $&$   0.01 $&$  1.5 $&$  0.01 $&$  1230 $&$  1840 $&$   0.25 \pm   0.02 $&$   1.95 \pm   0.14 $&$  32.0 \pm   5.3 $ \\
$   0.01 $&$   0.01 $&$  2   $&$  0.02 $&$  1450 $&$  3400 $&$   0.19 \pm   0.01 $&$   1.69 \pm   0.08 $&$  15.9 \pm   1.5 $ \\
$   0.01 $&$   0.01 $&$  2.5 $&$  0.03 $&$  1050 $&$  1745 $&$   0.13 \pm   0.01 $&$   1.38 \pm   0.07 $&$   7.39 \pm   1.02 $ \\
$   0.01 $&$   0.01 $&$  3   $&$  0.04 $&$   900 $&$  2200 $&$   0.12 \pm   0.02 $&$   1.31 \pm   0.09 $&$   5.27 \pm   1.11 $ \\
$   0.01 $&$   0.01 $&$  4   $&$  0.06 $&$  1150 $&$  2911 $&$   0.12 \pm   0.02 $&$   1.27 \pm   0.08 $&$   3.91 \pm   0.71 $ \\
$   0.01 $&$   0.01 $&$ 10   $&$  0.18 $&$  3990 $&$  5650 $&$   0.13 \pm   0.01 $&$   1.07 \pm   0.03 $&$   1.35 \pm   0.14 $ \\
\tableline
\end{tabular}
\caption{ Summary of the results. The first three columns are the run
  parameters, corresponding to those presented in Table 1. The fourth
  column shows the stability parameter $r$ defined in
  (\ref{eq:littler}). The 4th and 5th columns show the start and end
  times for the temporal average, as discussed in Section
  \ref{sec:gammaextract}. The 6th, 7th and 8th columns show the
  flux ratio $\gamma_{\rm tot}^{-1}$ and Nusselt numbers ${\rm Nu}_T$ and ${\rm Nu}_\mu$, as defined in equations (\ref{eq:ntgammadef}) and (\ref{eq:numudef}). Three significant digits are shown for ${\rm Nu}_T$ and ${\rm Nu}_\mu$, and two for $\gamma_{\rm tot}^{-1}$. }
 \end{center}
\label{run-results}
\end{table}

\begin{table}
\begin{center}
\begin{tabular}{ccccccccc}
\\
\tableline
Pr       &$ \tau  $&$ R_0^{-1} $&$ r   $&$ t_{\rm start} $&$ t_{\rm
  end} $&$ \gamma_{\rm tot}^{-1} $&$ {\rm Nu}_T $&$ {\rm Nu}_\mu $ \\
\hline
\hline
$   0.3 $&$   0.1 $&$  1.1 $&$  0.04 $&$   220 $&$   300 $&$   0.55 \pm   0.03 $&$   9.35 \pm   2.09 $&$  46.7 \pm  12.8 $ \\
$   0.3 $&$   0.1 $&$  1.2 $&$  0.09 $&$   300 $&$   450 $&$   0.43 \pm   0.01 $&$   5.36 \pm   0.15 $&$  19.4 \pm   0.8 $ \\
$   0.3 $&$   0.1 $&$  1.4 $&$  0.18 $&$   300 $&$   400 $&$   0.32 \pm   0.02 $&$   2.87 \pm   0.06 $&$   6.58 \pm   0.52 $ \\
$   0.3 $&$   0.1 $&$  1.7 $&$  0.31 $&$   460 $&$  1200 $&$   0.26 \pm   0.01 $&$   1.78 \pm   0.05 $&$   2.78 \pm   0.05 $ \\
$   0.3 $&$   0.1 $&$  2   $&$  0.44 $&$   620 $&$   920 $&$   0.25 \pm   0.01 $&$   1.42 \pm   0.04 $&$   1.81 \pm   0.12 $ \\
\hline
$   0.1 $&$   0.3 $&$  1.1 $&$  0.06 $&$   480 $&$   620 $&$   0.68 \pm   0.01 $&$   4.52 \pm   0.46 $&$   9.29 \pm   1.10 $ \\
$   0.1 $&$   0.3 $&$  1.2 $&$  0.11 $&$   650 $&$  1100 $&$   0.61 \pm   0.02 $&$   2.80 \pm   0.21 $&$   4.79 \pm   0.49 $ \\
$   0.1 $&$   0.3 $&$  1.3 $&$  0.17 $&$   660 $&$  1650 $&$   0.56 \pm   0.01 $&$   1.92 \pm   0.07 $&$   2.76 \pm   0.14 $ \\
$   0.1 $&$   0.3 $&$  1.5 $&$  0.29 $&$   850 $&$  1180 $&$   0.57 \pm   0.01 $&$   1.62 \pm   0.05 $&$   2.05 \pm   0.08 $ \\
$   0.1 $&$   0.3 $&$  2   $&$  0.57 $&$  1500 $&$  2350 $&$   0.63 \pm   0.01 $&$   1.17 \pm   0.01 $&$   1.23 \pm   0.01 $ \\
\hline
$   0.3 $&$   0.03 $&$  1.1 $&$  0.03 $&$   225 $&$   325 $&$   0.57 \pm   0.02 $&$  19.3 \pm   1.9 $&$ 332 \pm  42 $ \\
$   0.3 $&$   0.03 $&$  1.25$&$  0.09 $&$   190 $&$   500 $&$   0.36 \pm   0.06 $&$   6.62 \pm   2.26 $&$  63.3 \pm  33.4 $ \\
$   0.3 $&$   0.03 $&$  1.5 $&$  0.17 $&$   220 $&$   930 $&$   0.20 \pm   0.02 $&$   2.98 \pm   0.43 $&$  13.0 \pm   3.1 $ \\
$   0.3 $&$   0.03 $&$  2   $&$  0.34 $&$   270 $&$   937 $&$   0.12 \pm   0.01 $&$   1.64 \pm   0.10 $&$   3.17 \pm   0.35 $ \\
\hline
$   0.03 $&$   0.3 $&$  1.1 $&$  0.05 $&$   900 $&$  1420 $&$   0.72 \pm   0.01 $&$   4.63 \pm   0.32 $&$  10.1 \pm   0.9 $ \\
$   0.03 $&$   0.3 $&$  1.2 $&$  0.09 $&$  1100 $&$  1850 $&$   0.68 \pm   0.05 $&$   3.42 \pm   1.36 $&$   6.48 \pm   3.10 $ \\
$   0.03 $&$   0.3 $&$  1.35$&$  0.17 $&$  1300 $&$  4100 $&$   0.57 \pm   0.01 $&$   1.70 \pm   0.04 $&$   2.41 \pm   0.07 $ \\
$   0.03 $&$   0.3 $&$  1.5 $&$  0.24 $&$  1300 $&$  2077 $&$   0.57 \pm   0.01 $&$   1.50 \pm   0.04 $&$   1.91 \pm   0.07 $ \\
\tableline
\end{tabular}
\caption{(Continued from Table 6) }
 \end{center}
\end{table}

\newpage

\bibliographystyle{apj}

\end{document}